\documentclass[twocolumn,showpacs,aps,prd,superscriptaddress]{revtex4}

\usepackage{dcolumn}
\usepackage{amsmath}
\usepackage{graphicx}
\def\mpp  {\ensuremath{m_{p\overline{p}}}\xspace}
\def\mpks  {\ensuremath{m_{p\KS}}\xspace}

\def\bppkz {\ensuremath {B^0\to p\overline{p}K^0}\xspace}

\def\bppk {\ensuremath {B^+\to p\overline{p}K^+}\xspace}
\def\bpph {\ensuremath {B\to p\overline{p}h}\xspace}
\def\bppks {\ensuremath {B^0\to p\overline{p}\KS}\xspace}
\def\bpppi {\ensuremath {B^+\to p\overline{p}\pi^+}\xspace}

\def\bppkst {\ensuremath {B^0\to p\overline{p}K^{*0}}\xspace}
\def\bppkstp {\ensuremath {B^+\to p\overline{p}K^{*+}}\xspace}
\def\pph {\ensuremath {p\overline{p}h}\xspace}
\def\ppkz {\ensuremath {p\overline{p}K^0}\xspace}
\def\ppk {\ensuremath {p\overline{p}K^+}\xspace}

\def\pppi {\ensuremath {p\overline{p}\pi^+}\xspace}

\def\ppkst {\ensuremath {p\overline{p}K^{*0}}\xspace}
\def\ppkstp {\ensuremath {p\overline{p}K^{*+}}\xspace}

\newcommand{\BABARPubYear}    {07}

\newcommand{\BABARPubNumber}  {031}

\newcommand{\SLACPubNumber} {12661}

\input babarsym

\newcolumntype{.}{D{.}{.}{-1}}
\newcolumntype{p}{D{m}{\pm}{-1}}
\newcolumntype{+}{D{m}{\,\pm\,}{-1}}

\long\def\inst#1{\par\nobreak\kern 4pt\nobreak

    {\it #1}\par\vskip 10pt plus 3pt minus 3pt}

\begin{document}

{\pagestyle{empty}
\begin{flushleft}
\babar-PUB-\BABARPubYear/\BABARPubNumber \\
SLAC-PUB-\SLACPubNumber \\
%hep-ex/\LANLNumber \\
%\today \\
\end{flushleft}

%\par\vskip 3cm

% Title of the paper
\title{\Large \bf Evidence for the {\boldmath $B^0\rightarrow p\overline{p}K^{*0}$} and 
{\boldmath $B^+\to \eta_cK^{*+}$} decays and 
Study of the Decay Dynamics of {\boldmath $B$} Meson Decays into {\boldmath $p\overline{p}h$} 
Final States.
}
%\bigskip

% Input author list file
%% author list as of 04-May-2007 (570 authors)
%
\author{B.~Aubert}
\author{M.~Bona}
\author{D.~Boutigny}
\author{Y.~Karyotakis}
\author{J.~P.~Lees}
\author{V.~Poireau}
\author{X.~Prudent}
\author{V.~Tisserand}
\author{A.~Zghiche}
\affiliation{Laboratoire de Physique des Particules, IN2P3/CNRS et Universit\'e de Savoie, F-74941 Annecy-Le-Vieux, France }
\author{J.~Garra~Tico}
\author{E.~Grauges}
\affiliation{Universitat de Barcelona, Facultat de Fisica, Departament ECM, E-08028 Barcelona, Spain }
% ...
\collaboration{The \babar\ Collaboration}
\noaffiliation

% Abstract

\begin{abstract}
With a sample of 232$\times10^{6}$ $\Upsilon(4S)\rightarrow B\overline{B}$ events 
collected with the \babar\ detector, 
we study the decays of $B$ mesons to $p \overline{p} h$ 
final states, where $h=\pi^+,\,\KS,\,K^{*0}$ or $K^{*+}$. 
We report evidence for the \bppkst %B->ppbarK*0
decay, with a branching fraction (1.5$\pm$0.5(stat)$\pm$0.4(syst))$\times 10^{-6}$, 
and for the $B^+\rightarrow \eta_cK^{*+}$ decay, with the branching fraction of 
${\cal B}(B^+\rightarrow\eta_cK^{*+})\times{\cal B}(\eta_c\to p\overline{p})=$
($1.57^{+0.56}_{-0.45}$(stat)$^{+0.46}_{-0.36}$(syst))$\times 10^{-6}$,
and provide improved measurements of the branching fractions of the other modes of this type.
We also report the measurements of the charge asymmetry  consistent with zero  
in the \bpppi, \bppkst and \bppkstp modes.
No evidence is found for the pentaquark candidate $\Theta^{+}$
in the mass range 1.52 to 1.55\gevcc, decaying into $p\KS$, 
or the glueball candidate $f_J(2220)$ in 
the mass range $2.2<\mpp<2.4$\gevcc, 
and branching fraction limits are established for both at 
the $10^{-7}$ level. 

\vfill
\end{abstract}

% PACS, the Physics and Astronomy Classification Scheme.
 
\maketitle

}

% The body of the paper starts here

%\input{introduction}
\section{Introduction}

This paper reports measurements of the branching fractions of the baryonic three-body decays 
$B\rightarrow p\overline{p}h$ where $h=\pi^+,\,\KS,\,K^{*0}$, or $K^{*+}$ 
and their resonant substructures. 
This work is a continuation of our study of the \bppk decay reported in Ref.~\cite{babar}. 
Observations of these decays were reported earlier in Ref.~\cite{belle}, except for
the channel \bppkst, for which only an upper limit was obtained.

These channels are interesting for the dynamical information in the distribution of the three final-state particles
and for the possible presence of exotic intermediate states, such as
a pentaquark candidate $\Theta^+(1540)$ in the $m_{p\KS}$ spectrum~\cite{pent0} or
an $f_{J}(2220)$ glueball candidate in the \mpp spectrum~\cite{glue}.

The quark diagrams of the three-body baryonic \B decays are described in detail in Ref.~\cite{dia}; 
only the dominant contributions are described below.
In particular, the \bpppi decay proceeds mainly through 
external and internal $W-$emission ``tree'' processes, while in the decay 
$B^0\rightarrow p\overline{p}K^{0(*)}$    
a virtual loop ``penguin'' process $b\to s g$ is expected to be dominant. 
The $B^+\rightarrow p\overline{p}K^{+(*)}$ decay receives contributions 
from both the penguin and the doubly Cabibbo-Kobayashi-Maskawa-suppressed 
external $W-$emission tree processes.

An important feature of \bpph decays is an enhancement at low $p\overline{p}$ masses 
reported in Ref.~\cite{belle}, similar to that seen in several 
other baryonic decays of $B$ mesons~\cite{bb} and $J/\psi$~\cite{BES}. 
This enhancement might reflect short-range correlations between $p$
and $\overline{p}$ in a fragmentation chain~\cite{hou,dia}.
Alternatively, it could be a feature 
of a quasi-two-body decay in which the $p\overline{p}$ system is produced through 
an intermediate resonance~\cite{glue}, in particular the $X(1835)$ state 
recently observed by BES~\cite{BESX}.
Rosner~\cite{Rosner} proposed distinguishing between the two mechanisms by 
studying the distribution of events in the Dalitz plot.
As in the case of the resonance, $p$ and $\overline{p}$ are produced independent from a 
hadron, then the distributions $m_{ph}$ and $m_{\overline{p}h}$ should be
identical. 

Ten experimental groups reported narrow enhancements near 1.54\gevcc in the invariant-mass 
spectra for $nK^+$ and $pK^0$~\cite{pent}. The minimal 
quark content of a state that decays strongly to $nK^+$ is $dduu\overline{s}$: therefore, these mass peaks 
were interpreted as a possible pentaquark state, called $\Theta(1540)^+$. On the other hand, a number of experiments that 
observe strange baryons with mass similar to that of the $\Theta(1540)^+$ [e.g., $\Lambda(1520)\to pK^-$ ] see no evidence 
for the $\Theta(1540)^+$~\cite{PDG,nopent}.
As suggested in Ref.~\cite{pent0} we search for a pentaquark baryon candidate $\Theta^+$, 
in the \mpks mass distribution of \bppkz decays. 

A narrow state $f_J(2220)$ with a width of 23\mev was reported in the $K\overline{K}$ spectrum 
by the MARK III experiment~\cite{mark3} and later seen in $K^+K^-,\,K^0\overline{K}^0,\,\pi^+\pi^-$ 
and $p\overline{p}$ by the BES~\cite{besglue} experiment. 
However, its non-observation in quite a few $p\overline{p}$ annihilation modes, in particular 
in $p\overline{p}\to f_J(2220)\to K\overline{K}$~\cite{LEAR}, has led to doubts about 
its existence~\cite{close}. One theoretical conjecture~\cite{glue} suggests a possible presence of the $f_J(2220)$ 
resonance in baryonic $B$ decays. 

Direct \CP violation is observable as an asymmetry 
in yields between a decay and its \CP conjugate when at least two 
contributing amplitudes carry different weak and strong phases.  
Following the observation of the direct \CP violation in %penguin-dominated 
$B^0\rightarrow K^+\pi^{-}$~\cite{dircp} transitions and its non-observation in $B^+\rightarrow K^+\pi^{0}$~\cite{pi0},
it is interesting to study the charge asymmetry in the $B\rightarrow p\overline{p}h$ system.
Three-body baryonic \B decays to $p\overline{p}h$ occur through two different mechanisms 
(penguin and tree), which carry different weak phases and, in general, different strong phases. 
Although negligible direct \CP violation asymmetry is measured in the \bppk mode~\cite{babar,belle},
recent theoretical calculations predict a significant asymmetry in the \bppkstp mode~\cite{cpkstar}.

We use events with the same final state particles to isolate 
the decay $B^0\to \Lambda^+_c\overline{p}$, $\Lambda^+_c\to pK^{0(*)}$ and 
decays $B\rightarrow X_{c\overline{c}}h$, for $X_{c\overline{c}}=\eta_c$ 
and $J/\psi$, with a subsequent decay to $p\overline{p}$. 
Charge-conjugate reactions are included implicitly throughout the paper.

In Sec. II, we describe the \babar\ detector. In Secs. III, IV, and V we 
discuss the reconstruction of the \B-meson candidates,  
the branching fraction (B.F.) extraction and the systematic uncertainties, respectively. 
In Sec. VI we present our results on the \bpph, $B\rightarrow X_{c\overline{c}}h$ and 
$B^0\to \Lambda^+_c\overline{p}$ branching fraction measurements, 
charge-asymmetry measurements, and searches for $\Theta(1540)^+$ pentaquark 
and $f_J(2220)$ glueball states. We discuss the \bpph decay dynamics 
and summarize in Secs. VII and VIII, respectively.

\section{\babar\ Detector and Data Sample}

This measurement is performed with a data sample of 
232$\times10^{6}$ $\Upsilon(4S)\rightarrow B\overline{B}$ decays 
corresponding to an integrated luminosity of $210\,\mbox{fb}\,^{-1}$, 
collected at the $\Upsilon$(4S) resonance (``on-resonance'')
with the \babar\ detector at the PEP-II $e^+ e^-$ storage ring.
An additional $21\,\mbox{fb}^{-1}$ of data (``off-resonance''), 
collected about $40\,\mbox{MeV}$ below the resonance, is used to study the backgrounds
from light-quark and $c\overline{c}$ production.
The \babar\ detector is described in detail elsewhere \cite{nim}.
Charged-particle trajectories are measured by a five-layer silicon vertex tracker (SVT)
 and a 40-layer drift chamber (DCH), both operating in a 1.5-T solenoidal magnetic field. 
A Cherenkov radiation detector (DIRC) is used for charged-particle identification.
The CsI(Tl) electromagnetic calorimeter detects photon and electron showers.
The \babar\ detector Monte Carlo simulation based on GEANT4~\cite{simu} is used 
to optimize selection criteria and to determine selection efficiencies.

\section{{\boldmath $B$} meson reconstruction}

The $B^+$ and $B^0$ decays are reconstructed in the following modes: 
\bpppi, \bppkst, \bppks and \bppkstp. 
Charged tracks not coming from \KS must originate from the interaction point, 
have transverse momentum greater than 0.1\gevc, and at least twelve DCH hits.
Charged pions, kaons and protons are identified by the average energy loss ($dE/dx$)
in the tracking devices and by the pattern of Cherenkov photons in the DIRC.
The $\KS\to \pi^+ \pi^-$ candidates are required to have a two-pion invariant mass 
within 8\mevcc of the nominal \KS mass~\cite{PDG} and a cosine of the angle between the line connecting the 
$B$ and \KS decay vertices and the \KS momentum greater than 0.999 in the laboratory frame.
$K^{*+}$ candidates are reconstructed in the $K^{*+}\rightarrow K^{0}_{S}\pi^{+}$ 
decay channel, and $K^{*0}$ candidates are reconstructed in the  $K^{*0}\rightarrow K^{+}\pi^{-}$ mode.
Candidates whose mass is within 80\mevcc of the nominal $K^{*}$ mass~\cite{PDG} are selected as $K^{*}$ 
candidates, and the ones with mass $160-240$\mevcc away from the $K^{*}$ mass 
are chosen as the $K^*$ sideband sample. The three daughters in the $B$ decay are 
fit constraining their paths to a common vertex.
To suppress background, we impose a cut requiring a probability greater than 
$10^{-4}$ on the each of the \KS and the $B$ candidates' vertices. 

Two kinematic constraints of $B$-meson pair-production at the $\Upsilon(4S)$ 
are used to characterize the $B$ candidates:
the beam-energy-substituted mass 
$m_{ES}=[(E^2_{cm}/2+\mbox{\bf p}_0\cdot \mbox{\bf p}_B)^2/E^2_0-{\mbox{\bf p}^2_B}]^{1/2}$ 
and the energy difference $\Delta E=E^*_B-E_{cm}/2$, 
where $E_{cm}$ is the total center-of-mass energy, 
$E^*_B$ is the center-of-mass energy of the \B meson candidate 
(the asterisk denotes the center-of-mass frame throughout the paper), and 
$(E_0,\mbox{\bf p}_0)$ is the four-momentum of the initial state and 
$\mbox{\bf p}_B$ is the \B momentum in the laboratory frame. 
For signal \B decays the $\Delta E$ distribution peaks near zero with a  
mode-dependent resolution of $11-17\,$\mev, while the $m_{ES}$ distribution peaks near the \B meson mass 
with a mode-dependent resolution of $2.5-3.0\,$MeV/$c^2$, as determined in Monte Carlo simulation.

We select events with $\mes >5.22$\gevcc ($5.25$\gevcc for the \bpppi mode) and $|\DeltaE|<0.10$\gev 
($0.15$\gev for the \bppks mode). We make these selections quite loose since 
these two variables are used in a maximum likelihood fit to extract the signal yield. 
Of the candidates passing that loose selection, 
only one candidate is chosen for each event, selecting the one with 
the highest $B$ vertex probability, or, if a \KS is present in the decay chain, the highest
value of the product of the $B$ and the \KS vertex probabilities.
To improve the resulting mass resolutions, after selecting the $B$ candidates 
we perform a kinematic fit fixing the mass of each
$B$ candidate to its known value~\cite{PDG} and its energy to a half of the 
total center-of-mass energy.

The background is dominated by random combinations of tracks created in $e^+e^-\rightarrow q\overline{q}$ 
(where $q$ is $u,\,d,\,s,c$ quarks) continuum events. 
These events are collimated along the original quark directions and can be distinguished from 
more spherical $B\overline{B}$ events. 
We construct a Fisher discriminant~(${\cal F}$)~\cite{fisher} as  
a linear combination of the following four event-shape variables: 
\begin{enumerate}
\item cos$\theta^*_{thr}$, the cosine of the angle between the thrust axis of the reconstructed $B$ and the 
beam axis; 
\item cos$\theta^*_{mom}$, the cosine of the angle between 
the momentum of the reconstructed $B$ and the beam axis; 
\item and 4. the zeroth and the second Legendre polynomial momentum moments, $L_0$ and $L_2$.
They are defined as follows: $L_0=\sum_{i} |{\bf p}^*_i|$ and $L_2=\sum_{i} |{\bf p}^*_i|[(3\cos^2\theta^*_{thr_{B,i}}-1)/2]$, 
where ${\bf p}^*_i$ are the momenta of the tracks 
and neutral clusters not associated with the $B$ candidate and $\theta^*_{thr_{B,i}}$ is
the angle between  ${\bf p}^*_i$ and the thrust axis of the $B$ candidate.
\end{enumerate}
The coefficients for the Fisher discriminant variables  
are determined for each of the modes separately 
by maximizing the separation between the means of the signal and background 
Fisher discriminant distributions obtained from the signal Monte Carlo samples and the 
off-resonance data events, respectively.

There is also a background for the three-body decays \bpph  
from $B$ mesons decaying into the same final states as the signal, such as 
$B\rightarrow X_{c\overline{c}}h$ decays, where $X_{c\overline{c}}=\eta_c,\,J/\psi,\,\psi',\chi_{c0},\chi_{c1},\chi_{c2}$ 
decaying to $p\overline{p}$ and the $B^0\to \Lambda^+_c\overline{p}$ decay. 
This background, which is of interest in its own right, is discussed in the 
next section. Other backgrounds from \B decays are negligible. 
 
\section{Branching Fraction Extraction}\label{sec-bfex}

We perform an unbinned extended maximum likelihood (ML) fit to extract the signal yields. The variables \mes, 
\DeltaE and ${\cal F}$ are used as discriminating variables to separate signal from background.

To estimate the contribution from \B decays that proceed through $b\to c$ transitions leading to the \pph
final state, such as  $\eta_c,\,J/\psi \to p\overline{p}$ 
and $\Lambda^+_c\to p\KS/K^{*0}$, the maximum likelihood fit is performed in 
three distinct regions:
\begin{enumerate}

\item The main ``charmonium'' region which includes the $\eta_c$ and $J/\psi$ resonances:
$2.85<$\mpp$<3.15\,$GeV/c$^2$ (for $h=K^0/K^{*0}/K^{*+}$). 

\item The ``charm''($\Lambda^+_c$) region delimited by 
$|m_{ph}-2.3\gevcc|<0.1\,$GeV/c$^2$ where $h=\KS/K^{*0}$, and  
$|\mpp-3\gevcc|>0.15\,$GeV/c$^2$ (this prevents overlap with the charmonium region).
Note that for $h=\KS$ both $m_{ph}$ and $m_{\overline{p}h}$ combinations are considered.

\item The ``all-other'' region. Significant background from slow pions in the \bpppi mode 
leads to a slight difference of the Fisher discriminant shape at low and high \mpp. 
To reduce the sensitivity to that correlation we perform the fit for this mode in 
two regions: \mpp$<3.6$\gevcc and \mpp$>3.6$\gevcc. 
\end{enumerate}

In the charmonium region, in addition to the three variables described above, 
the \mpp variable is used to distinguish between the non-resonant signal and 
contributions from $\eta_c$ and $J/\psi$. 
In the charm region the corresponding additional variable is $m_{ph}$. 

The data sample is assumed to consist of two components: signal events, 
including $B$ meson decays to $X_{c\overline{c}}h$ and $\Lambda^+_c\overline{p}$, which have 
the same final state particles as the signal, 
and combinatorial background events.
The latter are due to random combinations of tracks from both continuum and $B\overline{B}$ events.
For the fit in the charmonium region, $\eta_c$ and $J/\psi$ signal-like components are included in the fit,
a $\Lambda^+_c$ component corresponding to the overlap of charm and charmonium regions is not
included here but its contribution is taken into account in Sec.~\ref{sec-charmlessbf}. 
For the charm region, a $\Lambda^+_c$ component is included in the fit as both signal and background contributions.
For the decay \bpppi, an additional \bppk component is considered, to account for 
the \bppk events present in the fit region because of kaon/pion mis-identification. 
The signal component is split into correctly reconstructed true events and 
mis-reconstructed events, so-called self-cross-feed (SCF).
The SCF events are signal events in which one or more
of the tracks from the signal side are lost and a track from the other $B$ decay is included in the 
reconstruction. The fraction of SCF events is determined with a \bpph Monte Carlo sample and 
varies from 0.5$\%$ for the \bppks mode to $5.6\%$ for the 
\bppkstp mode. In the \bppks mode, the SCF signal events are indistinguishable from the 
combinatorial background and no separate SCF signal component is used in the maximum likelihood fit. 

In the maximum likelihood fit, each component is modeled by the product of  
probability density functions (PDF) of the following variables, which are assumed to be uncorrelated for all components: 
\begin{itemize}
\item \mes, \DeltaE, ${\cal F}$ and \mpp in the charmonium region:
\begin{equation}
\label{eq:pdf1}
{\mathcal P}_1^x = {\mathcal P}^x(\mes,\DeltaE,{\mathcal F},\mpp)
\end{equation}
\item \mes, \DeltaE and ${\cal F}$ and $m_{ph}$ in the charm region:
\begin{equation}
{\mathcal P}_2^x = {\mathcal P}^x(\mes,\DeltaE,{\mathcal F},m_{ph})
\end{equation}
\item \mes, \DeltaE and ${\cal F}$ in the all-other region:
\begin{equation}
\label{eq:pdf3}
{\mathcal P}_3^x = {\mathcal P}^x(\mes,\DeltaE,{\mathcal F})
\end{equation}
\end{itemize} 
where $x$ indicates the event source.  
In the all-other region there are two components: $x$ is either signal ($s$) or background ($b$),
in the charmonium region there are four components $x=s,\,\eta_c,\,J/\psi$ and $b$; and
in the charm region there are three components $x=s,\,\Lambda^+_c$ and $b$.

The likelihood is given by 
\begin{equation}
{\mathcal L}_r=e^{-N_r}\prod^{N_r}_{i=1} \sum_{x} N_r^{x}{\mathcal P}_r^x ,
\end{equation}
where $r$ corresponds to a fit region, $N_r$ is the total number of events in that region 
and $\underset{x}\sum$ is the sum over all the corresponding fit components in a given region. 
Then the total number of events in the charmonium region
is  $N_1=N_1^{s}+N^{\eta_c}+N^{J/\psi}+N_1^{b}$, 
in the charm region is $N_2=N_2^{s}+N^{\Lambda^+_c}+N_2^{b}$, 
and  in the all-other region is $N_3=N_3^{s}+N_3^{b}$, where
$N^{s}$, $N^{\eta_c}$, $N^{J/\psi}$, $N^{\Lambda^+_c}$ and $N^{b}$ are 
the number of non-resonant signal, $\eta_c$, $J/\psi$, $\Lambda^+_c$ signal and combinatorial background events, 
respectively. ${\mathcal P}^{x}$ is the PDF for a corresponding component $x$, e.g. ${\mathcal P}^{b}_r$
is the background PDF for region $r$.  
The signal PDF ${\mathcal P}^{s}_r={\mathcal P}^\text{true}_r+f_\text{SCF}{\mathcal P}^\text{SCF}_r$ 
consists of the PDFs of the true and SCF signal events, respectively,
with the corresponding fraction of SCF events $f_\text{SCF}$ as 
determined from \bpph signal Monte Carlo and fixed in the fit. 

\begin{table*}
\caption{The PDF parametrization of $B\to p\overline p h$ signal, self-cross-feed (SCF), 
$\eta_c$, $J/\psi$, $\Lambda^+_c$ and combinatorial 
background contributions. We use the following notation: 
``BifGauss'' for a Bifurcated Gaussian, ``G'' for a Gaussian 
($2\times$G are two Gaussian distributions with common mean), ``Voigtian''  for a convolution of 
a Breit-Wigner distribution and a sum of Gaussian distributions with common mean, 
``ARGUS'' for a threshold function~\cite{Argus}, ``norm'' for normalization 
and ``ratio'' is a ratio of the normalizations of a linear to 
a Gaussian contributions. The parameters  floated for each of the functions are specified in the brackets.}
\label{modpdfpar}
\begin{center}
\begin{tabular}{llllll}
\hline
{Component} & \mes & \DeltaE & $\mathcal{F}$ & \mpp & $m_{ph}$ \\\hline\hline
\multicolumn{6}{c}{\bppks}\\\hline
{signal} & BifGauss (mean) & G$+$G (narrow mean) & G$+$G & constant (norm) & constant (norm) \\
{$J/\psi$} & {same as signal}&{same as signal}&{same as signal} & $2\times$G (narrow $\sigma$)  & $-$ \\
{$\eta_c$} & {same as signal}&{same as signal}&{same as signal} & $2\times$Voigtian  & $-$ \\
{$\Lambda_c^+$} & {same as signal}&{same as signal}&{same as signal} & $-$ & $2\times$G \\
{combinatorial} & ARGUS (slope) & linear (all) & G$+$G (means) & linear & linear$+2\times$G (ratio) \\
\hline\hline
\multicolumn{6}{c}{\bppkstp}\\\hline
{signal} & BifGauss (mean) & G$+$G (narrow mean) & G$+$G & constant (norm) & $-$ \\
{$J/\psi$} & {same as signal}&{same as signal}&{same as signal} & $2\times$G (narrow $\sigma$) & $-$ \\
{$\eta_c$} & {same as signal}&{same as signal}&{same as signal} & $2\times$Voigtian  & $-$ \\
\text{SCF} & ARGUS & linear & G$+$G & constant & $-$ \\
{combinatorial} & ARGUS (slope) & linear (all) & G (all) & linear  & $-$ \\
\hline\hline
\multicolumn{6}{c}{\bppkst}\\\hline
{signal} & BifGauss (mean) & G$+$G (narrow mean) & G$+$G & constant (norm) & constant (norm) \\
{$J/\psi$} & {same as signal}&{same as signal}&{same as signal} & $2\times$G (narrow $\sigma$) & $-$ \\
{$\eta_c$} & {same as signal}&{same as signal}&{same as signal} & $2\times$Voigtian  & $-$ \\
{$\Lambda_c^+$} & {same as signal}&{same as signal}&{same as signal} & $-$ & $2\times$G \\
\text{SCF} & ARGUS$+$G & quadratic & G$+$G & constant & $-$ \\
{combinatorial} & ARGUS (slope) & linear (all) & G (all) & linear & linear$+2\times$G (ratio) \\
\hline\hline
\multicolumn{6}{c}{\bpppi}\\\hline
{signal} & BifGauss (mean) & G$+$G (narrow mean) & G$+$G & $-$ & $-$ \\
\text{SCF} & ARGUS & linear & G$+$G & $-$ & $-$ \\
{$p\overline{p}K^+$} & BifGauss & G$+$G & G$+$G & $-$ & $-$ \\
{$p\overline{p}K^+$ SCF} & ARGUS & linear & G$+$G & $-$ & $-$ \\
{combinatorial} & ARGUS (slope) & linear (all) & G$+$G (all) & $-$ & $-$\\
\hline
\end{tabular}
\end{center}
\end{table*}

The parametrization of the $B\to p\overline p h$ true signal, SCF, 
$\eta_c$, $J/\psi$, $\Lambda^+_c$ and combinatorial 
background components is summarized in Table~\ref{modpdfpar}.

All parameters in the \bpph true and SCF signal 
PDFs are obtained from corresponding \bpph signal Monte Carlo simulations.
The \bpph signal Monte Carlo events are 
simulated according to a three-body phase-space decay expectation, which has been 
shown in Refs.~\cite{belle,babar} not to reproduce the data. In order to improve the parametrization 
for the \bpph signal PDFs the signal Monte Carlo events are re-weighted according to the 
\mpp distributions from \bppk given in Refs.~\cite{belle,babar}.

The \mes, \DeltaE, and $\cal{F}$ PDFs for the $\eta_c$, $J/\psi$ and $\Lambda^+_c$ 
contributions are taken to be the same as the corresponding PDFs for the \bpph signal.
The known masses of the $J/\psi$ and $\Lambda_c^+$ and 
the known mass and width of the $\eta_c$ are used in the fits.
All the other parameters that are not 
floated are fixed to the values obtained from the corresponding Monte Carlo simulation.

All the parameters for the \bppk background events in the \bpppi mode 
are fixed in the fit to the values 
obtained from the \bppk Monte Carlo events re-weighted as described above 
as well as the expected amount of the charmonium events with \ppk final states.

For the combinatorial background events  all the parameters fixed in the fit 
are obtained from the on-$\Upsilon(4S)$ resonance data \mes sideband ($\mes<5.26$\gevcc).

The parametrization of the signal and combinatorial background PDFs does not vary with the fit region.

The remaining floating parameters of the PDFs are determined by a maximum likelihood fit to the data.
The fit uses the kinematic variables described above but is independent of the
 location of the event in the Dalitz plot. This fit not only determines the
 various parameters of the PDFs, but also the number of signal and
 background events, and the covariance matrix for these event numbers.

Once the maximum likelihood fit provides the best estimate of the PDF parameters, 
we use a weighting technique~\cite{splot} to measure the branching fraction and 
reconstruct efficiency-corrected mass distributions. This method allows us to take into 
account the dependence of the efficiency on the position of a candidate in the Dalitz plot.
We assume that the distributions in the variables \mes, \DeltaE, and $\cal{F}$ 
 are uncorrelated with the location in the Dalitz plot.
 
The event-dependent weight to be signal, $\mathcal{W}^j_{ \mathrm{s} }$
(also known as the {\it sWeight} in the {\it sPlots} method~\cite{splot}), is 
computed from the PDFs:

\begin{equation}
\mathcal{W}^j_{\mathrm{s}}=\frac{\sum_{x} {\bf V}_r(s,x){\mathcal P}_r^{x,j}}
{\sum_{x} N^{ }_r{\mathcal P}^{x,j}_r},
\end{equation}
where ${\mathcal P}^{x,j}_r$ is the value of the PDF for the component $x$ 
($x\,=$ signal, background, charmonium-signal, charmonium-$\eta_c$, etc.) in 
the event $j$ for the fit region $r$ (as defined in Equations~\ref{eq:pdf1}-\ref{eq:pdf3}), and 
${\bf V}_r(s,x)$ is the covariance between the number of signal events $N_r^{s}$ and 
the  number of events for the component $x$, $N_r^{x}$, in the fit region $r$ defined by:
\begin{equation}
{\bf V}_r(s,x)^{-1}=\frac{\partial^2(-{\ln\mathcal L})}{\partial N_r^{s}\partial N_r^{x}}=% \\
\sum^{N_r}_{j=1}\frac{{\mathcal P}_r^{s,j}{\mathcal P}_r^{x,j}}{(\sum_{x} N_r^{x}{\mathcal P}_r^{x,j})^2}.%\\
\end{equation}
The sum of $\mathcal{W}^j_{ \mathrm{s} }$ over all events $j$ is just $N^s=N_1^s+N_2^s+N_3^s$, 
while the sum of $\mathcal{W}^j_{ \mathrm{s} }$ over all events in a small area in phase space gives, 
in the limit of high statistics, the correct distribution 
of the signal in that area while preserving the total signal yield. 
These weights ($\mathcal{W}^j_{ \mathrm{s} }$) allow optimal discrimination between signal-like and background-like events~\cite{splot} and serve 
as a tool to reconstruct the resulting signal distributions. 

The resulting branching fraction is then calculated as follows:
\begin{equation}
\label{eq:BF}
\BR = \sum_{j=1}^N \frac{ \mathcal{W}^j_{ \mathrm{s} }  }{N_{\BB}\cdot \varepsilon_j \cdot \BR_\text{sub}},
\end{equation}
where the sum is over all events $j$, ${N_{\BB}}$ is the total number of \BB pairs, 
$\BR_\text{sub}$ is the product of the branching fractions in the sub-decays and 
$\varepsilon_j$ is the reconstruction efficiency of the event $j$. The reconstruction efficiency depends on the 
position of the candidate on the Dalitz plane and is obtained from the corresponding signal Monte Carlo simulations. 
The statistical error on the branching fraction is given by 
\begin{equation}
\label{eq:BFerr}
\frac{\Delta\BR}{\BR} =\frac{\sqrt{\sum_{j=1}^N \frac{ (\mathcal{W}^j_{ \mathrm{s} })^2  }{\varepsilon^2_j}}}{\sum_{j=1}^N \frac{ \mathcal{W}^j_{ \mathrm{s} }  }{ \varepsilon_j}}.
\end{equation}

\section{Systematic errors}

\begin{table*}
\caption{Systematic errors (in percent) and efficiency corrections for all modes. 
The efficiency correction is a weight applied event-by-event to signal efficiency 
to account for residual differences between signal Monte Carlo and data.} 
\label{all-tab:systematics}
\begin{center}
\begin{tabular}{lccccccccccc}
\hline\hline
Error Source                             & \ppkz  & \ppkstp   & \ppkst    & \pppi       &$\eta_cK^{0}$  &$\eta_cK^{*+}$&$\eta_cK^{*0}$&$\overline{p}\Lambda_c^+(pK^{0})$ &$\overline{p}\Lambda_c^+(pK^{*0})$&$\Theta^+$ \\\hline
\BB counting                             & 1.1     & 1.1        & 1.1        & 1.1      & 1.1             & 1.1        & 1.1          & 1.1          & 1.1  & 1.1           \\      
PID efficiency                           & 1.2 & 1.6       &    3.4  & 2.1      & 1.2             & 1.6        & 2.9          & 1.6          & 3.6  & -5.8/+6.6   \\           
Track reconstruction                     & 1.6     & 2.4        & 3.6        & 2.4      & 1.6             & 2.4        & 3.2          & 1.6          & 3.2  & 1.6        \\             
$K^0_S$ reconstruction                       & 2.8     & 2.9  & n/a        & n/a      & 1.1             & 1.2        & n/a          & 2.1          & n/a  & 1.6        \\             
Monte Carlo Statistics                   & 2.0     & -3.4/+4.4  & 2.5  & 2.6 & 0.3             & 1.3        & 0.8          & 1.2          & 1.0  & 0.3       \\                   
Dalitz plot binning                      & 2.1     & 3.5        & 1.4        & 2.0      &      n/a           &  n/a          &    n/a          &     n/a         &  n/a    &    n/a          \\
Pre-selection                            & 0.5     & 0.8        & 0.4        & 0.3      & 0.5             & 0.8        & 0.4          & 0.5          & 0.4  & 0.5       \\              
Fit Bias                                 & 5.0     & 3.0        & 3.6        & 1.1      & 3.4             & 3.4        & 1.5          & 20.0         & 11.0 &          n/a   \\             
PDF Parametrization                     & 3.0     & -3.7/+3.2  & -1.9/+2.2  & 6.2      & 1.1             &-3.3/+3.8   & 3.0    & 1.3    & 0.6  &     n/a       \\             
Fit Region                               & 5.7     & 17.0       & 6.5        & 13.0        & 5.0             & 1.8        & 3.0          & 5.0          & 3.4  &       n/a     \\              
\B Bkg / B.F. errors                      & 1.0     & 0.8        & 2.9        & 0.8      &   n/a              &    n/a        &     n/a         &     n/a         &   n/a      &  n/a       \\
Non-resonant $K\pi$ & n/a     & 15.4 & 25.2  & n/a & n/a             &-21.9/+22.7   & 15.5          & n/a         & -42.7/+46.9 & n/a   \\\hline        
Total($\%$)                                    & 9.5     & 24.4 & 27.1  & 15.2 & 6.7             &-22.8/+23.6   & 16.8          & 21.0         & -44.5/+48.6 & -6.3/+7.1   \\\hline             
Pre-selection correction                     & 0.976   & 0.967      & 0.968      & 0.983    & 0.976           & 0.967      & 0.968        & 0.969        & 0.968& 0.976      \\             
\KS correction                           & 0.981 & 0.980 & n/a  &  n/a                                   & 0.972           & 0.966      & n/a          & 0.976        & n/a     & n/a  \\\hline  
\end{tabular}
\end{center}
\end{table*}

\begin{table*}
\caption{Systematic errors for the charge asymmetry measurements.} 
\label{all-tab:sysas}
\begin{center}
\begin{tabular}{l.......}
\hline\hline
Error   & \multicolumn{1}{r}{\ppkstp}   & \multicolumn{1}{r}{\ppkst}    & \multicolumn{1}{r}{\pppi}    & \multicolumn{1}{r}{$\eta_cK^{*+}$}&\multicolumn{1}{r}{$\eta_cK^{*0}$} & \multicolumn{1}{r}{$J/\psi K^{*+}$}&\multicolumn{1}{r}{$J/\psi K^{*0}$}\\\hline
Track Reconstruction  & 0.002 & 0.004 & 0.002 & 0.002 & 0.004 & 0.002 & 0.004 \\ 
PID Efficiency        & 0.003 & 0.015 & 0.001 & 0.001 & 0.002 & 0.002 & 0.001 \\  
Monte Carlo Statistics         & 0.012 & 0.057 & 0.021 & 0.017 & 0.004 & 0.002 & 0.016 \\
Fitting               & 0.050 & 0.020 & 0.040 & 0.066 & 0.040 & 0.018 & 0.007 \\\hline
Total                 & 0.05 & 0.06 & 0.05 & 0.07 & 0.04 & 0.02 & 0.02 \\\hline
\end{tabular}
\end{center}
\end{table*}

The contributions to the systematic errors on the measured branching fractions and charge asymmetries are summarized in Tables~\ref{all-tab:systematics} 
and \ref{all-tab:sysas}. 

The $\Upsilon(4S)$ is assumed to decay equally to $B^0\overline{B^0}$ and $B^+B^-$ mesons. 
Incomplete knowledge of the luminosity and cross-section leads to a $1.1\%$ uncertainty in all branching fraction 
measurements. 
Charged-tracking, particle-identification (PID) and \KS reconstruction studies of the data lead to small 
corrections applied to each track in these simulations. Limitations of statistics and purity in these data-Monte 
Carlo comparisons lead to residual systematic uncertainties. 
Limitation of Monte Carlo statistics employed in each analysis contributes a small uncertainty. 
The effects of binning on the efficiency are estimated by studying the variation in the resulting signal 
yield due to changes in the chosen bin size. 

A large control sample of $B$$\rightarrow$$J/\psi(e^+e^-,\mu^+\mu^-)h$ is 
studied separately in data and Monte Carlo simulations 
to understand uncertainties arising from the choice of vertexing cuts and from the 
parametrization of the PDFs for $\Delta E$, $m_{ES}$ and ${\cal F}$ (this uncertainty is labeled ``Pre-Selection''
in Table~\ref{all-tab:systematics}).

The uncertainty due to possible correlation between the fit variables is estimated by performing fits to a number of Monte Carlo 
experiments that consist of fully simulated signal events embedded in parametrized background 
simulations. This so-called ``Fit Bias'' uncertainty in Table~\ref{all-tab:systematics}) is on the order of a few percent. 

Where the Monte Carlo values are used to fix signal shape parameters in a fit, we obtain the uncertainty on the signal yield 
due to a change in each of the PDF parameters 
by doing a number of toy Monte Carlo experiments with parametrized signal and background distributions.
In each a fit is performed with the original value of the 
PDF parameter and the value shifted by its uncertainty, 
which is obtained from the \bpph signal Monte Carlo simulation.
The resulting uncertainties are added together  in quadrature,  taking into account 
the correlation between the PDF parameters, to obtain the total error for each PDF.
The procedure is repeated using the correlation matrices between the variables    
to obtain the total error for each fit component, 
which is then added in quadrature to give the total error on PDF parametrization.

In a similar fashion, different fit ranges for charmonium and charm regions 
are employed and the resulting variation of the fit yields is taken as a systematic error. 

The potential correlation of the fit variables with their location on the 
Dalitz plot would slightly violate the main principle of the $sPlot$ method.
To reduce the sensitivity to that correlation 
the fit is performed in one, two and four \mpp regions for all the modes. 
The resulting variation of the branching fraction is taken as a systematic error.
Note that we constrained the \bppkstp fit to make the signal component 
non-negative (in the charmonium region), which results in additional systematic error on the fit region. 
All the uncertainties above are summed in quadrature to give the ``Fit Region'' error in Table~\ref{all-tab:systematics}).

Branching fraction uncertainties~\cite{PDG} on 
${\cal B}(B^+\rightarrow Xh)\times{\cal B}(X\rightarrow p\overline{p})$, where $X=\chi_{c[0,1,2]},\psi'$ and $h=\KS,\,K^{*0}$ or $K^{*+}$, 
and on the sub-branching fractions of $K^0$ and $K^{*}$, which affect the \bpph branching fraction measurements, 
are given in the ``\B Bkg / B.F. errors'' line in Table~\ref{all-tab:systematics}.

For modes that contain $K^*$ mesons, the non-resonant $K\pi$ background is obtained by
performing a maximum likelihood fit
and extracting the branching fraction in the $K^{*}$ sideband region.
The non-resonant $K\pi$ contributions to the
branching fraction values, based on the $K^*$ sideband data,
 for the the \bppkstp and \bppkst modes are
(0.34$\pm$0.74)$\times10^{-6}$ and (0.23$\pm$0.30)$\times10^{-6}$, 
correspondingly. The $K^*$-sideband signal yields 
for $J/\psi K^{*0}$ and $J/\psi K^{*+}$
are $11.5^{+4.3}_{-3.5}$ and $1.2^{+1.7}_{-0.9}$ events; for 
$\eta_cK^{*0}$ and $\eta_cK^{*+}$ 
are $0\pm5.7$ and $1.8^{+2.9}_{-2.0}$ events; for the $B\to\Lambda^+_c\overline{p}$
in the \bppkst mode $4.5^{+3.6}_{-2.7}$ events. As no significant non-resonant $K\pi$ 
background is seen, we do not perform the 
sideband subtraction, but rather add these background contributions and 
their statistical uncertainties in quadrature to give the ``Non-resonant $K\pi$'' systematic
uncertainties listed in Table~\ref{all-tab:systematics}.

All the uncertainties described above are added in quadrature in the ``Total'' line in Table~\ref{all-tab:systematics}. 
The final branching fraction value obtained from Eq.~\ref{eq:BF} is scaled by 
Pre-selection and ``\KS'' corrections. 

\section{Results}

\subsection{{\boldmath $B\to p\overline{p}h$} Branching Fraction Measurements}
\label{sec-charmlessbf}

\begin{figure}
\begin{center}
\includegraphics[width=8.4cm]{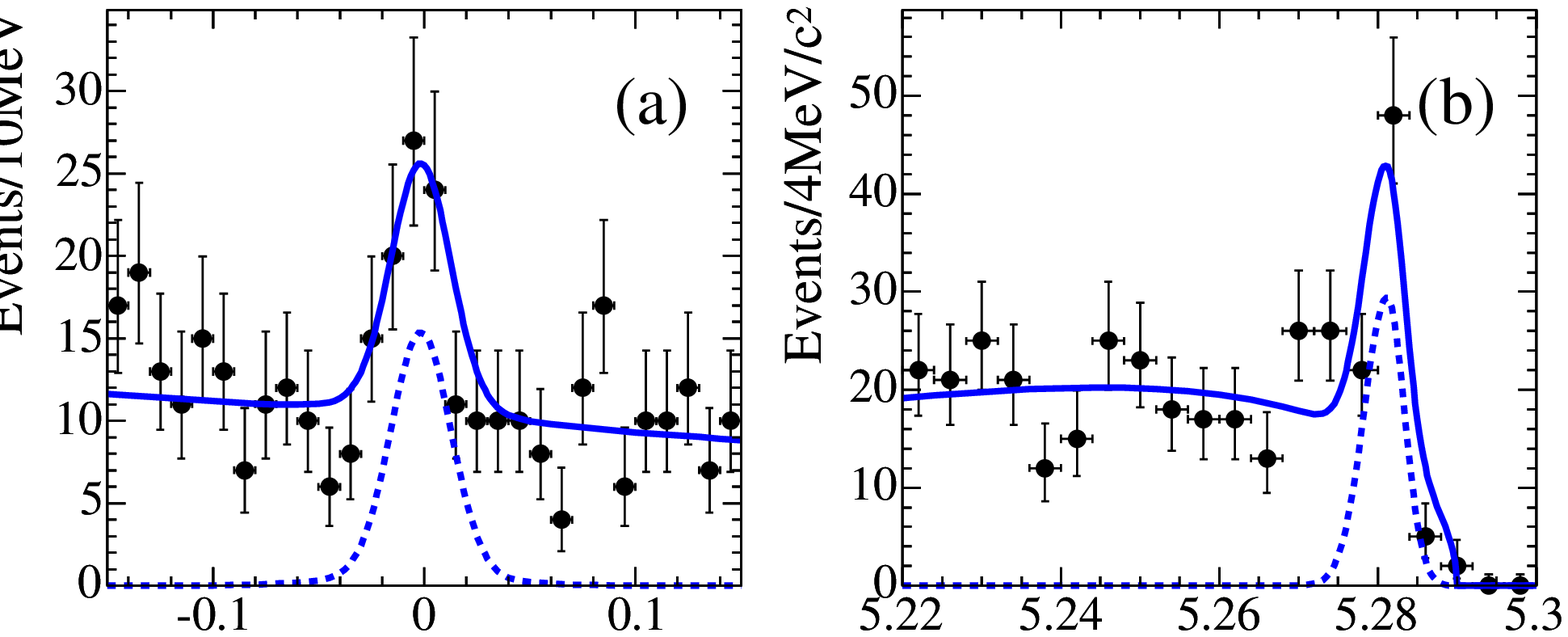}  
\includegraphics[width=8.4cm]{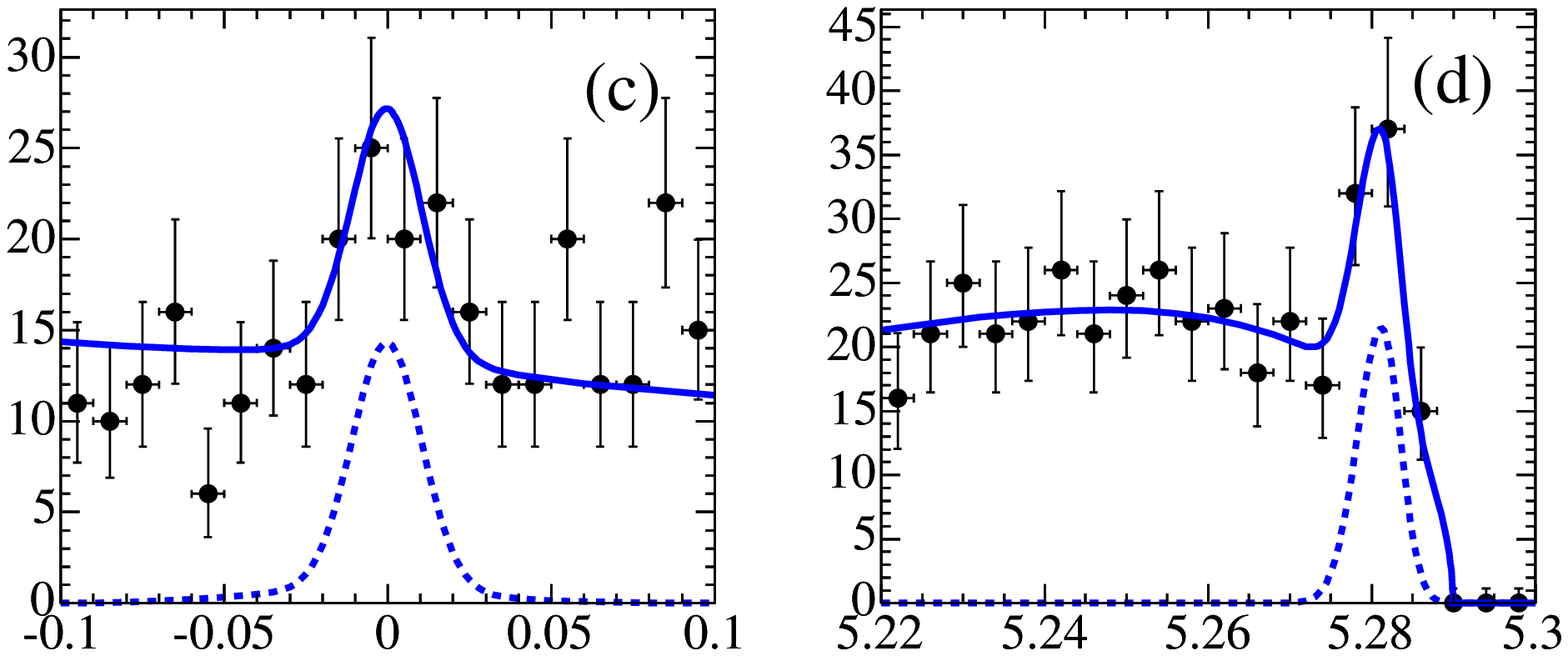}  
\includegraphics[width=8.4cm]{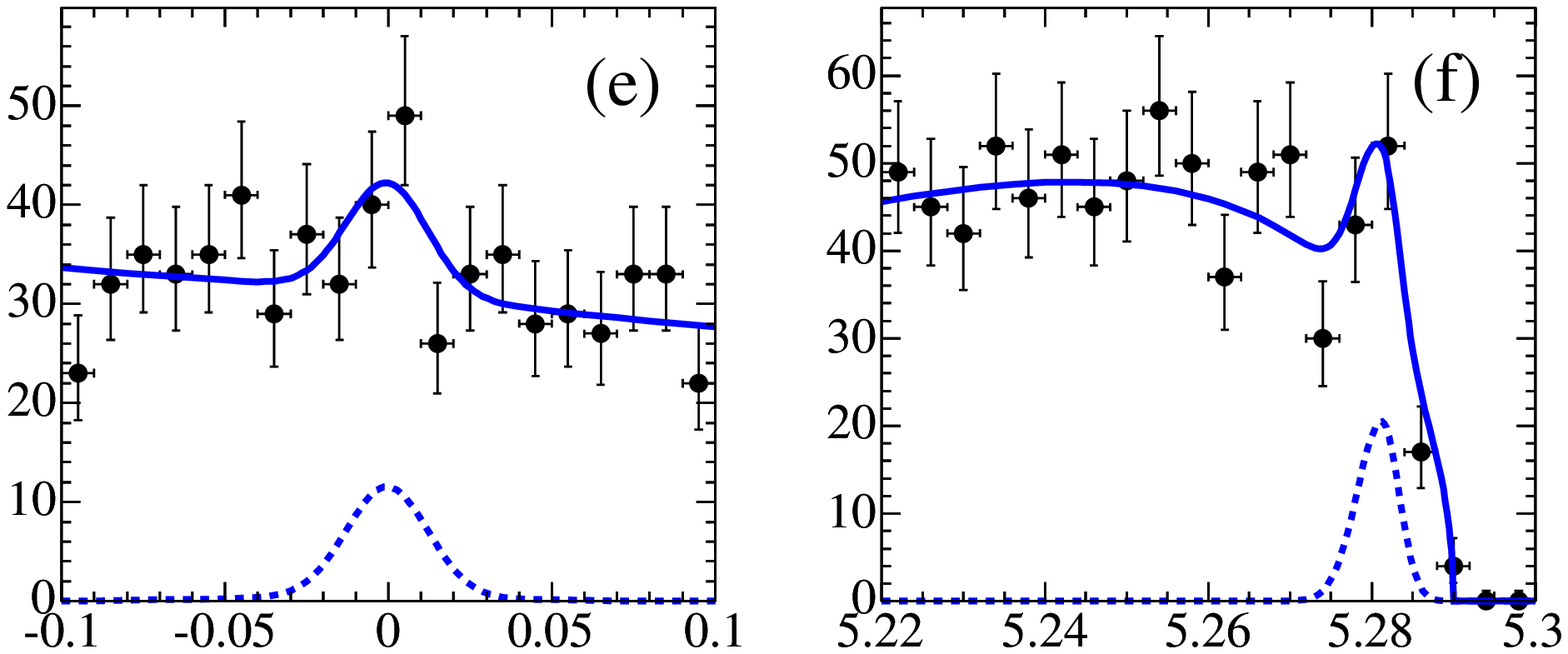}  
\includegraphics[width=8.4cm]{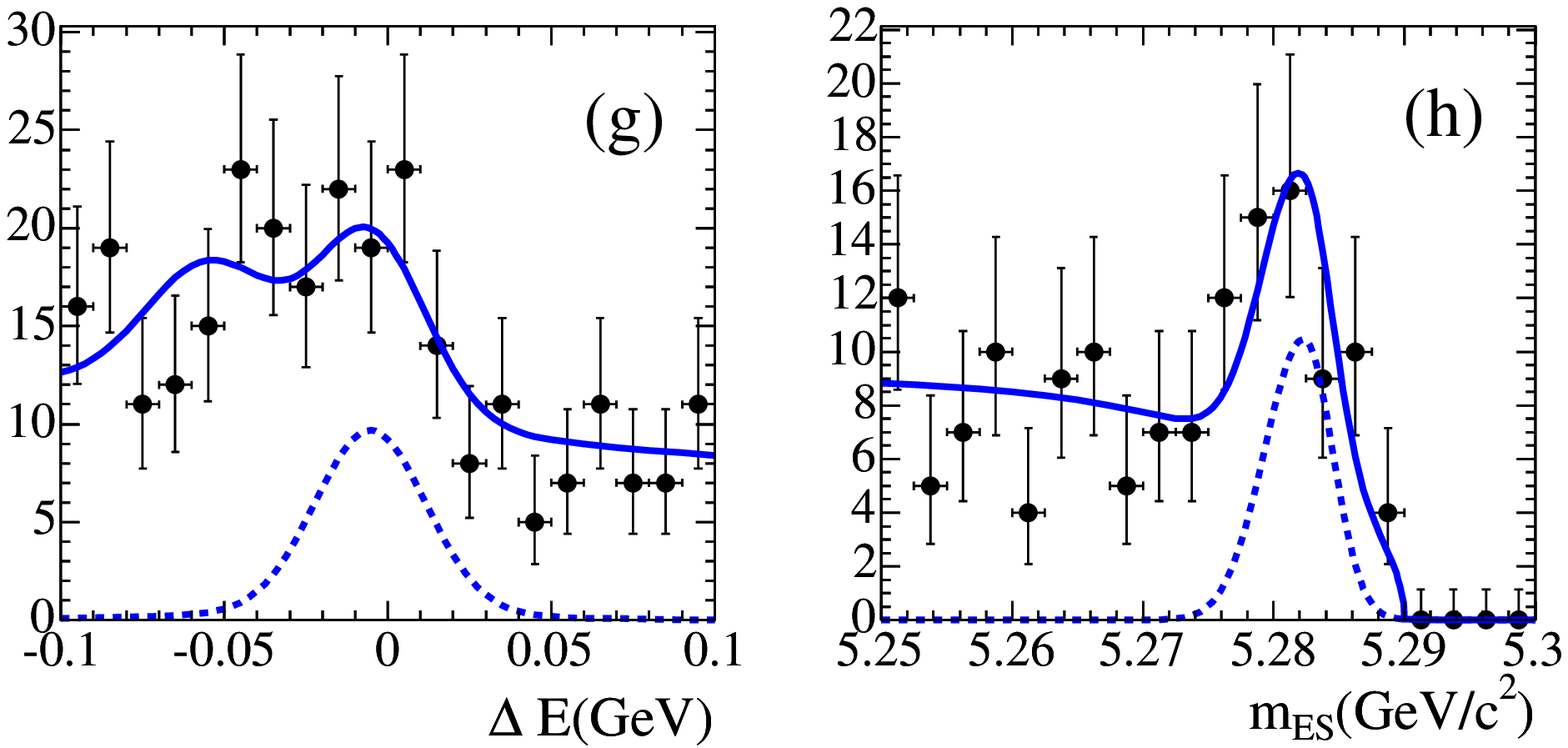}  
\caption{Distributions (points with error bars) of \DeltaE  and \mes for 
data candidates passing loose selection in the  
 \bppkz (a)-(b), \bppkstp (c)-(d), \bppkst (e)-(f), \bpppi (g)-(h) modes.
The superimposed  solid  curves represent
the projections on the three-dimensional-fit PDFs on the respective axis,
and the dashed line is the fitted signal contribution.
Note that the excess of events on the left of the signal peak in (g) corresponds to the 
\bppk final states, where $K^+$ is mis-identified as $\pi^+$.
The following loose selection is applied: 
(a) $\mes>5.275$\gevcc and $\mathcal{F}>0.2$;
(b) $|\DeltaE|<0.02$\gevcc and $\mathcal{F}>0.2$;
(c) $\mes>5.275$\gevcc and $\mathcal{F}>0.2$;
(d) $|\DeltaE|<0.02$\gevcc and $\mathcal{F}>0.2$;
(e) $\mes>5.275$\gevcc and $\mathcal{F}>0.6$;
(f) $|\DeltaE|<0.015$\gevcc and $\mathcal{F}>0.6$;
(g) $\mes>5.275$\gevcc and $\mathcal{F}>1.5$;
(h) $|\DeltaE|<0.015$\gevcc and $\mathcal{F}>1.5$.
} 
\label{splotRes1}
\end{center}
\end{figure}

\begin{table}
\caption{Summary of the resulting yields from the maximum likelihood fit for all modes.}
\center
\setlength{\extrarowheight}{2pt}
\begin{tabular}{lrrrr}
\hline\hline
{Region} & {charmonium} &{charm}& {All-Other} & {Total}\\\hline
{Component} & \multicolumn{4}{c}{\bppks mode.} \\\hline
{Signal} & {$17^{+9}_{-8}$} & {$3^{+4}_{-3}$} & {70$^{+12}_{-11}$}&{90$^{+16}_{-14}$} \\
{$\eta_c$} & {23$^{+8}_{-7}$} & & & {23$^{+8}_{-7}$}\\
{$J/\psi$} & {53$^{+8}_{-7}$} & & & {53$^{+8}_{-7}$}\\
{$\Lambda^+_c$} & & {6.8$^{+3.6}_{-2.8}$} & & {6.8$^{+3.6}_{-2.8}$}\\
{Background} & {1152$\pm$34} &{1096$\pm33$}&{14769$\pm$122}&{17017$\pm$131}\\\hline

{} & \multicolumn{4}{c}{\bppkstp mode.} \\\hline
{Signal} & {$0\pm9$} &  & {52$^{+11}_{-10}$}&{52$\pm 14$} \\
{$\eta_c$} & {12.3$^{+4.4}_{-3.6}$} & & & {12.3$^{+4.4}_{-3.6}$} \\
{$J/\psi$} & {34$\pm 6$} & & &{34$\pm 6$}\\
{Background} & {766$\pm$28} &&{10063$\pm$101}&{10829$\pm$105}\\\hline

{} & \multicolumn{4}{c}{\bppkst mode.} \\\hline
{Signal} & {8$^{+10}_{-8}$} & {$4^{+6}_{-4}$} & {50$^{+14}_{-13}$}&{63$^{+18}_{-16}$} \\
{$\eta_c$} & {37$^{+10}_{-9}$} & & & {37$^{+10}_{-9}$}\\
{$J/\psi$} & {106$\pm 11$} & & &{106$\pm 11$}\\
{$\Lambda^+_c$} & & {12.3$^{+4.9}_{-4.1}$} & & {12.3$^{+4.9}_{-4.1}$}\\
{Background} & {2207$\pm$47} &{1971$\pm45$}&{26312$\pm$163}&{30490$\pm$176}\\\hline

{} & \multicolumn{4}{c}{\bpppi mode.} \\\hline
{Signal} & & & {185$\pm$28}&{185$\pm$28} \\
{$p\overline{p}K^+$} & & & {157$\pm$30}&{157$\pm$30} \\
{Background} &  &&{90438$\pm$305}&{90438$\pm$305}\\\hline
\end{tabular}
\label{mlfitres}
\end{table}

The event yields from the maximum likelihood fit are presented in Table~\ref{mlfitres}, while Fig.~\ref{splotRes1} shows projections of the 
corresponding three-dimensional PDFs on the \DeltaE and \mes axes. 
The measured branching fractions calculated from Eq.~\ref{eq:BF} (scaled by Pre-selection and \KS corrections)
are summarized in Table~\ref{bfres}, where the effective efficiency for the modes observed is calculated accounting for the 
correct event distribution in the Dalitz plot. 

Our fitting method removes all $B\to\eta_c h$ and $B\to J/\psi h$ (except in the \bpppi mode) contributions 
and most of the $B^0\to \Lambda^+_c \overline{p}$ contributions. 
There is still some remaining $B^0\to \Lambda^+_c \overline{p}$  background contribution from the 
$B^0\to \Lambda^+_c \overline{p}$ events in the charmonium region.  Knowing the relative efficiencies of
$B^0\to \Lambda^+_c \overline{p}$ Monte Carlo events inside and outside the charmonium region allows us to 
calculate the remaining $B^0\to \Lambda^+_c \overline{p}$ background contribution to be 
$(0.04\pm0.02)\times 10^{-6}$ and $(0.06\pm0.03)\times 10^{-6}$ for the \bppkz and the \bppkst modes,
respectively.

The remaining unknown background comes from the $B\to \chi_{c0} h$ events. 
It is possible to estimate the $B^0\rightarrow\chi_{c0}K^0$ branching fraction using the corresponding 
$B^+\rightarrow\chi_{c0}K^+$ branching fraction measurement. 
Because of isospin symmetry one would expect ratios of the 
charged and neutral $B$ mesons decaying into $\chi_{c0}$ and $\chi_{c1}$ to be equal.
Thus we estimate~\cite{PDG}: ${\cal B}(B^0\rightarrow\chi_{c0}K^0)\approx {\cal B}(B^+\rightarrow\chi_{c0}K^+)\frac{{\cal B}(B^0\rightarrow\chi_{c1}K^0)}{{\cal B}(B^+\rightarrow\chi_{c1}K^+)}=(1.6^{+0.5}_{-0.4})\times 10^{-4}\times \frac{3.9\pm0.4}{5.3\pm0.7}=(0.12\pm0.04)\times 10^{-3}$.
This number needs to be multiplied by the ${\cal B}(\chi_{c0}\rightarrow p\overline{p})=(0.22\pm0.03)\times10^{-3}$  
and results in the expected contributions to the branching
fraction from this mode of $(0.026\pm0.009)\times 10^{-6}$.
The resulting contribution to the absolute systematic error on the \B background
is $0.009\times 10^{-6}$.
The contribution from the $B\to \chi_{c0}h$ events is ignored for other modes~\cite{chic0other}.
 
From  Ref.~\cite{PDG} we estimate the contribution of  $B^+\to J/\psi\pi^+$ to the \bpppi mode and 
the remaining charmonium contributions ($B\to\chi_{c1}h$ and $B\to\psi(2S)h$) for all the other modes. 
Since only an upper limit exists for the branching fraction of the $B\to\chi_{c2}h$
mode, it is not subtracted but taken as a one sided contribution the to systematic uncertainty.
The total expected $B$ background contribution is quoted in the line ``$B$ bkgr. B.F.'' 

The line ``Final B.F.'' in Table~\ref{bfres} summarizes the values of the charmless and charmoniumless \bpph  
branching fractions after $B$ background subtraction.

\begin{table*}
\caption{Summary of the 
non-charm, non-charmonium \bpph branching fraction calculation including $B$ background subtraction. 
Note that effective efficiency includes the sub-decay branching fractions as well as the correct 
event distribution in the Dalitz plot.}
\center
\begin{tabular}{lrrrr}
\hline\hline
{}& \ppkz & {\ppkstp}&{\ppkst}&{\pppi} \\\hline
{Sum of Signal Weights}                          &{181$\pm$29}          &{52$\pm$14}          &{63$\pm$17} & {185$\pm$28} \\
{Effective Efficiency,$\%$}                   &{24.5}            &{4.1}           &{16.0} &{44.6} \\
{B.F. from Eq.~\ref{eq:BF} ($10^{-6}$)}       &{3.17$\pm$0.53}           &{5.45$\pm$1.49}          &{1.70$\pm$0.45} &{1.79$\pm$0.29} \\
{$B$ bkgr. B.F. ($10^{-6}$)}                  & {0.22$\pm$0.03}          &{0.17$\pm$0.04}          & {0.23$\pm$0.05} & {0.10$\pm$0.01} \\\hline
{Final B.F. ($10^{-6}$)} & { 3.0$\pm$0.5$\pm$0.3}  & { 5.3$\pm$1.5$\pm$1.3} & { 1.5$\pm$0.5$\pm$0.4} &  {1.7$\pm$0.3$\pm$0.3}
 \\\hline
\end{tabular}
\label{bfres}
\end{table*}

We report the first evidence for the \bppkst decay 
with a significance of $3.2\sigma$ (including systematic uncertainties).
The statistical significance $\sigma$ throughout the paper is taken as  
$\sqrt{-2\ln ({\mathcal L}(0)/{\mathcal L}_\text{max})}$, where  ${\mathcal L}(0)$ is the 
likelihood of the fit assuming zero signal events and 
${\mathcal L}_{max}$ is the likelihood obtained in the full fit. 
For the \bppkst decay the ${\mathcal L}(0)$ is taken to be not at zero signal events, 
but at the expected number of the \B background events as discussed above.
To obtain the value of significance including the systematic uncertainties the 
likelihood function is smeared with a Gaussian distribution which has a width of the corresponding 
systematic  uncertainty.

The measurements of branching fractions for the $B\to p\overline{p}h$ modes from Ref.~\cite{belle} and this work are 
summarized in Table~\ref{sum-bf} and compared to those of the two-body mesonic modes from Refs.~\cite{PDG,HFAG,n1}.

The branching fractions are approximately two times smaller  
for the \bppkstp and the \bpppi modes, when compared to the Belle measurements~\cite{belle}, 
bringing the branching fraction of the \bppkstp mode below that of the \bppk mode   
and more in line with theoretical predictions~\cite{dia}.
However, the two experiments are in agreement within their errors.  

Since the virtual loop ``penguin'' process $b\to s g$ preserves
isospin we would naively expect the  ratio of the rates for \bppk and \bppkz to be 
unity as it is in two-body mesonic modes $B \to \rho^0h$ and $B\to\pi^0h$, 
but it is closer to two (see Table~\ref{sum-bf}). This could be explained by  
absence of the external $W-$emission Feynman tree diagram for the neutral $B$ mode. 
However, if this tree diagram were important, we would expect a 
much larger rate for \bpppi  than for \bppk, in contradiction with the data.
The \bppkstp branching fraction is also larger (by a factor of three) 
than that of \bppkst similar to the pattern suggested by the data for decays to $\rho^0K^*$ and $\pi^0K^*$.
The $\B\to p\overline{p}K^*$ modes are consistently smaller than the $\B\to p\overline{p}K$ modes 
in both the charged and neutral cases. 
This seems to be the case for the $\B\to\pi^0h$ modes as well, but not for the $\B\to\rho^0h$ modes.
The \bpppi branching fraction is lower than that of the \bppk mode as expected because the $b\to u$ transition 
at tree level is suppressed compared to the $b\to s$ penguin. This is similar to what is observed in the 
$\B\to\pi^0h$ modes but contrary to what is observed in $\B\to\rho^0h$.

Overall, the theoretical calculations of the baryonic $B$ decays are not very certain and 
the current measurements of the branching fractions of all four $\B\to p\overline{p}K$ modes 
are a further challenge to our understanding.

\begin{table*}
\begin{center}
\caption{Summary of the experimental values for the branching fractions ($\times10^{-6}$) of $B\rightarrow p\overline p h$
and their comparison to two-body mesonic modes. The values for the two-body mesonic modes are taken from Ref.~\cite{PDG}
unless otherwise noted. The values in {\bf bold} are those presented in the current work.}
\vspace{0.3cm}
\setlength{\extrarowheight}{2pt}
\begin{tabular}{crrrr}
\hline\hline
{$h$}     &{Belle $B\rightarrow p\overline{p}h$~\cite{belle}}                   & {\babar\ $B\rightarrow p\overline{p}h$}             & {$B\rightarrow\pi^0h$}   & {$B\rightarrow \rho^0h$}\\\hline 
{$K^+$}   &{$6.0\pm0.3\pm0.4$}     &{$6.7\pm0.5\pm0.4$\cite{babar}}                & $12.1\pm0.8$ & {$5.0^{+0.7}_{-0.8}$} \\
{$K^0$}   &{$2.08^{+0.52}_{-0.38}\pm\,0.24$}     &{\bf 3.0$\pm$0.5$\pm$0.3}           & $11.5\pm1.0$ & {$5.4^{+0.9}_{-1.0}$~\cite{HFAG}}\\
{$K^{*+}$}&{$10.3^{+3.6}_{-2.8}\,^{+1.3}_{-1.7}$}&{\bf 5.3$\pm$1.5$\pm$1.3}           &  $6.9\pm2.4$  & {$11.0\pm 4.0$}\\
{$K^{*0}$}&{$<7.6$, 90$\%$ CL}                   &{\bf 1.5$\pm$0.5$\pm$0.4}  & {$<3.5$, 90$\%$ CL}  & {$5.6\pm1.6$~\cite{n1}}\\
{$\pi^+$} &{$1.68^{+0.19}_{-0.17}\pm0.12$}     &{\bf 1.7$\pm$0.3$\pm$0.3}          & $5.5\pm0.6$  & {$8.7\pm1.1$}\\\hline
\end{tabular}
\label{sum-bf}
\end{center}
\end{table*}

\subsection{{\boldmath $B\to (J/\psi,\eta_c)h$} Branching Fraction Measurements}

\begin{figure}
\begin{center}
\includegraphics[height=5cm,width=8.5cm]{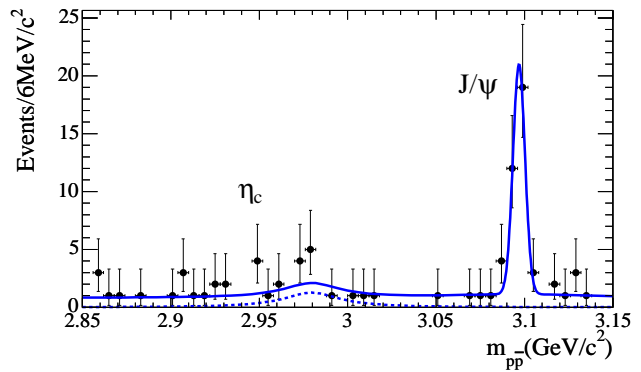}
\caption{The results of the maximum likelihood fit for \mes$>5.27$\gevcc 
and $|\DeltaE|<0.03$\gev for the \bppkstp mode. 
The solid line represents results of the fit and the dashed line shows the $\eta_c$ yield.}
\label{ml-cc}
\end{center}
\end{figure}

Using the $\eta_c$ and $J/\psi$ yields from Table~\ref{mlfitres}  
we obtain the branching fractions shown in Table~\ref{all-ccbf}. 
The values obtained are consistent with current world averages~\cite{PDG}. 
We also report the first evidence for the $B^+\to\eta_cK^{*+}$ decay, with a significance of 
3.2$\sigma$ (including systematic uncertainties). A sample of the maximum likelihood fit result 
for the charmonium region is shown in Fig.~\ref{ml-cc}.

\begin{table*}
\caption{Summary of the resulting branching fractions for the $\eta_c$ and $J/\psi$ modes 
(the order of the uncertainties is as follows: statistical, systematic, due to partial branching fraction correction where appropriate).
The following values of branching fractions are used ${\cal B}(\eta_c\to p\overline{p})=(1.3\pm0.4)\times10^{-3}$ and 
${\cal B}(J/\psi\to p\overline{p})=(2.12\pm0.10)\times10^{-3}$~\cite{PDG}.}
\center
\setlength{\extrarowheight}{2pt}
\begin{tabular}{ccrrrr}
\hline\hline
{$p\overline{p}X$} & {Efficiency} & \multicolumn{2}{c}{${\cal B}(B\rightarrow\eta_c(p\overline{p})X)\,(10^{-6})$} 
& \multicolumn{2}{c}{${\cal B}(B\to\eta_cX)\,(10^{-3})$} \\\hline
{Mode} & {$\%$} & {Measured} & {PDG~\cite{PDG}}  & {Measured} & {PDG~\cite{PDG}} \\\hline
{$p\overline{p}K^0$} & 36.3 & {$0.83^{+0.28}_{-0.26}\pm0.05$}& {1.56$\pm$0.71} & {$0.64^{+0.22}_{-0.20}\pm0.04^{+0.28}_{-0.15}$} & {1.2$\pm$0.4} \\
{$p\overline{p}K^{*0}$} & {23.7} & {$1.03^{+0.27}_{-0.24}\pm0.17$} & {2.08$\pm$1.11} & {$0.80^{+0.21}_{-0.19}\pm0.13^{+0.35}_{-0.19}$} 
& {1.6$\pm$0.7} \\
{$p\overline{p}K^{*+}$} & {15.7} & {$1.57^{+0.56+0.45}_{-0.46-0.36}$} & {-} & {$1.21^{+0.43+0.34+0.54}_{-0.35-0.28-0.28}$} 
& {-} \\\hline\hline
{Mode}& {Efficiency} & \multicolumn{2}{c}{${\cal B}(B\rightarrow J/\psi(p\overline{p}) X)\,(10^{-6})$} &  
\multicolumn{2}{c}{${\cal B}(B\to J/\psi X)\,(10^{-3})$} \\\hline
{$p\overline{p}K^0$} & {37.1} & {$1.87^{+0.28}_{-0.26}\pm0.07$} & {1.80$\pm$0.08} & {$0.88^{+0.13}_{-0.12}\pm0.03\pm0.04$} & {0.85$\pm$0.04} \\
{$p\overline{p}K^{*0}$} & {25.0} & {$2.82^{+0.30+0.36}_{-0.28-0.35}$} & {2.78$\pm$0.20} & {$1.33^{+0.14}_{-0.13}\pm0.17 ^{+0.07}_{-0.06}$} 
& {1.31$\pm$0.07} \\
{$p\overline{p}K^{*+}$} & {17.8} & {$3.78^{+0.72+0.28}_{-0.64-0.23}$} & {2.86$\pm$0.25} & {$1.78^{+0.34+0.13+0.09}_{-0.30-0.11-0.08}$} 
& {1.35$\pm$0.10} \\\hline
\end{tabular}
\label{all-ccbf}
\end{table*}

\subsection{{\boldmath $B^0\to \Lambda^+_c\overline{p}$} Branching Fraction Measurement}

\begin{table*}
\caption{Summary of the resulting branching fractions for $B^0\to \Lambda^+_c\overline{p}$ 
(the order of the uncertainties is as follows: 
statistical, systematic, the uncertainty in the ratio of 
the ${\cal B}(\Lambda^+_c\rightarrow pK^0/K^{*0})$ to the ${\cal B}(\Lambda^+_c\rightarrow pK\pi)$~\cite{PDG} and 
the uncertainty in the ${\cal B}(\Lambda^+_c\rightarrow pK\pi)$ value~\cite{PDG}). Note that 
efficiency does not include the sub-decay branching fractions.
}
\center
\setlength{\extrarowheight}{2pt}
\begin{tabular}{ccccc}
\hline\hline
{$\Lambda^+_c$ decay} & {Efficiency, $\%$} & {${\cal B}(\Lambda^+_c\rightarrow pX)(10^{-3})$\cite{PDG}} & {${\cal B}(B^0\rightarrow\Lambda^+_c\overline{p})(10^{-6})$} & {statistical significance} \\\hline
{$pK^0$} & {$25.4$} & {$23\pm6$} & $15.1^{+8.0}_{-6.2}\pm3.2^{+1.4+5.3}_{-1.2-3.2}$ & 3.4 \\
{$pK^{*0}$} & {$19.7$} & {$16\pm5$} & $26.9^{+10.7+13.0+4.0+9.4}_{-9.0-12.0-3.1-5.5}$ & 4.3 \\\hline
\end{tabular}
\label{all-lcbf}
\end{table*}

From the $B\to \Lambda^+_c\overline{p}$ fit yields in the \bppks and \bppkstp 
modes given in Table~\ref{mlfitres} 
we obtain the branching fractions shown in Table~\ref{all-lcbf}.
Averaging results for both modes and adding the errors in quadrature (except the systematic error on $B-$counting), 
we obtain the branching fraction  
${\cal B}(B^0\rightarrow\Lambda^+_c\overline{p})=(21.0^{+6.7}_{-5.5}$(stat)$^{+6.7}_{-6.2}$(syst)$
^{+2.1}_{-1.7}(\Lambda^+_c\,\,B.F.)
^{+7.4}_{-4.3}({\cal B}_{\Lambda^+_c\rightarrow pK\pi}))\times10^{-6}$.
This measurement is consistent with the current value of 
${\cal B}(B^0\rightarrow\Lambda^+_c\overline{p})=(21.9^{+5.6}_{-4.9}\pm3.2\pm5.7)\times10^{-6}$ 
based on a single measurement by Belle~\cite{2bodyex2}.

\subsection{{\boldmath $B\to p\overline{p}h$} Charge Asymmetry Measurements}\label{sec-chas}

The \CP-violating charge asymmetry is defined as $A_{ch}=(N_{\overline{B}}-N_{B})/(N_{\overline{B}}+N_{B})$, where 
$N_{B}$ and $N_{\overline{B}}$ are the event yields in each of the categories of interest.
The events yields are obtained from the maximum likelihood fit described in the Sec.~\ref{sec-bfex},
by integrating over the resulting signal event weights for each of the two charge categories separately. 
The resulting yields for all the modes (except \bppkz, which does not have information on the flavor of \B meson) 
are summarized in Table~\ref{all-chasym}. 
The measurements for the current modes are consistent with zero within less than three standard deviations.

\begin{table}
\caption{Summary of the asymmetry study: event yields for each of the categories of interest and the charge asymmetry.
We observe no statistically significant \CP asymmetries.}
\center
\begin{tabular}{lrrr}
\hline\hline
{} & {$N_{\overline{B}}$} &{$N_{B}$}& {$A_{ch}$}\\\hline
{Events Type} & \multicolumn{3}{c}{\bppkst mode.} \\\hline
{Signal} &                   {35$\pm$6}& {28$\pm$5} &{0.11$\pm$0.13$\pm$0.06} \\
{$\eta_c$} &                 {23$\pm$5}& {13$\pm$4} & {0.28$\pm$0.16$\pm$0.04}\\ 
{$J/\psi$} &                 {63$\pm$8}& {43$\pm$7} & {0.19$\pm$0.10$\pm$0.02}\\
{Background} &               {15050$\pm$123}& {15443$\pm$125} & {-0.013$\pm$0.006}\\\hline
{} & \multicolumn{3}{c}{\bppkstp mode.} \\\hline
{Signal} & {34$\pm$6} & {18$\pm$4} & {0.32$\pm$0.13$\pm$0.05} \\
{$\eta_c$} & {7$\pm$3} & {5$\pm$2} & {0.20$\pm$0.28$\pm$0.07} \\
{$J/\psi$} & {18$\pm$4} & {16$\pm$4} & {0.07$\pm$0.17$\pm$0.02} \\
{Background} &               {5453$\pm$74}& {5377$\pm$73} & {0.007$\pm$0.010}\\\hline\hline
{} & \multicolumn{3}{c}{\bpppi mode.} \\\hline
{Signal} & {97$\pm$10} & {89$\pm$9} & {0.04$\pm$0.07$\pm$0.04} \\
{Background} &               {45434$\pm$213}& {44968$\pm$212} & {0.005$\pm$0.003}\\\hline
\end{tabular}
\label{all-chasym}
\end{table}

\subsection{{\boldmath ${\cal B}(B^0\rightarrow\Theta(1540)^+\overline{p})$} Upper Limit Calculation}

As suggested in Ref.~\cite{pent0}, we search for a pentaquark baryon candidate, $\Theta^+$, in the \mpks mass distribution
of \bppkz decays. If $\Theta^+$ decays strongly, there are only two possible decays modes: $nK^+$ and $pK^0$. For this measurement we 
assume ${\cal B}(\Theta\rightarrow pK^0_S)=25\%$.
From dedicated signal Monte Carlo we determine that the $\Theta^+$ invariant mass resolution is 
represented by a sum of two Gaussian functions with common mean. The resolution 
of the main (secondary) Gaussian is 0.95(2.32)\mevcc and the wider Gaussian contributes $19\%$
of the total. The overall resolution, defined as the full width at half maximum of the resolution function divided by 2.4,
is 1\mevcc at the $\Theta^+$ mass of 1.54\gevcc. The $\Theta^+$ pentaquark signal efficiency is 
30.8$\pm$0.1$\%$. 
No events are observed in the $\Theta^+$ region of $1.52<m_{p\KS}<1.55$\gevcc (see Ref.~\cite{thesis} for details).
A Bayesian approach is used to calculate the upper limit of $0.20\times10^{-6}$ at 90$\%$ confidence level, 
assuming Poisson-distributed events in the absence of background, 
 including a multiplicative systematic error of $7.1\%$.
This value is  consistent with and improves on the upper limit from the Belle Collaboration ${\cal B}(B^0\rightarrow\Theta(1540)^+\overline{p})<0.92\times10^{-6}$~\cite{belle}.

\subsection{Search for glueball {\boldmath $f_J(2220)$} in {\boldmath $B\to p\overline{p}h$} decays}

We search for the narrow state $f_J(2220)$ by 
scanning through the $2.2<\mpp<2.4$\gevcc region with a $30$\mevcc mass  
window in the final states \ppkz and $p\overline{p}K^*$. This procedure is described in detail in Ref.~\cite{thesis}.
The largest upper limits at 90$\%$ confidence level, including multiplicative systematic uncertainties of 2.7(4.3 and 4.8)$\%$,  
on the product of branching fractions 
are found to be 
${\cal B}(B\to f_J(2220) h)\times {\cal B}(f_J(2220)\to p\overline{p})<4.5(7.7$ 
and $1.5)\times10^{-7}$ for $h=K^0(K^{*+}$ and $K^{*0})$, respectively, 
assuming the $f_J(2220)$ width is less than 30\mev.

\begin{figure}
\begin{center}
\includegraphics[width=9cm]{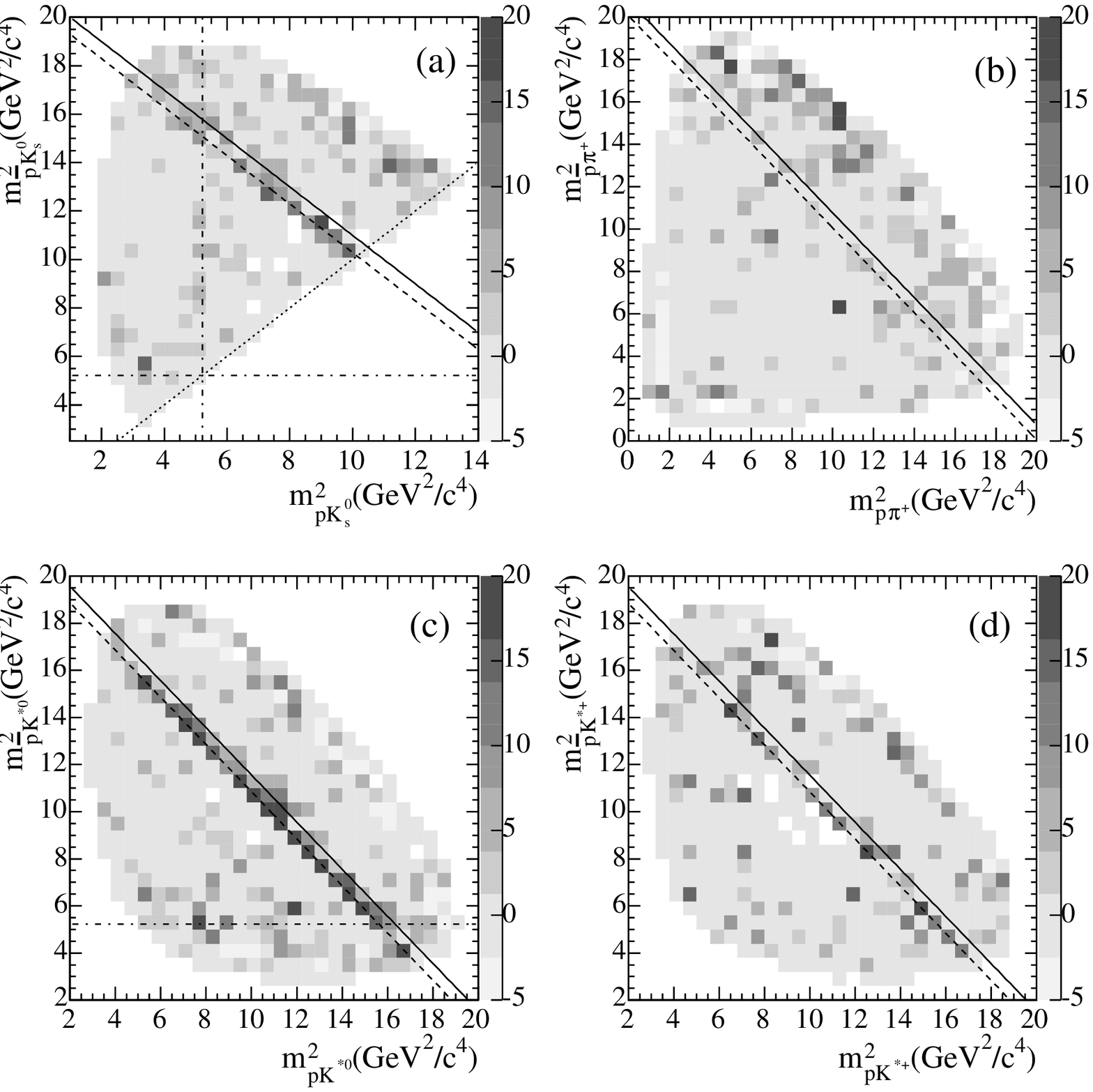}  
\caption{Dalitz plots for signal in the (a) \bppkz, (b) \bpppi, (c) \bppkst and (d) \bppkstp modes 
obtained using the weighting technique described in Ref.~\cite{splot}. The positions of the following 
resonances are shown: $\eta_c$ in solid, $J/\psi$ in dashed and $\Lambda_c$ in dot-dashed lines.
Note, that because of fluctuations and uncertainties, the signal rate in many bins is negative.
The white areas correspond to regions with no entries.}
\label{alldal}
\end{center}
\end{figure}

\begin{figure}
\begin{center}
\includegraphics[width=9cm]{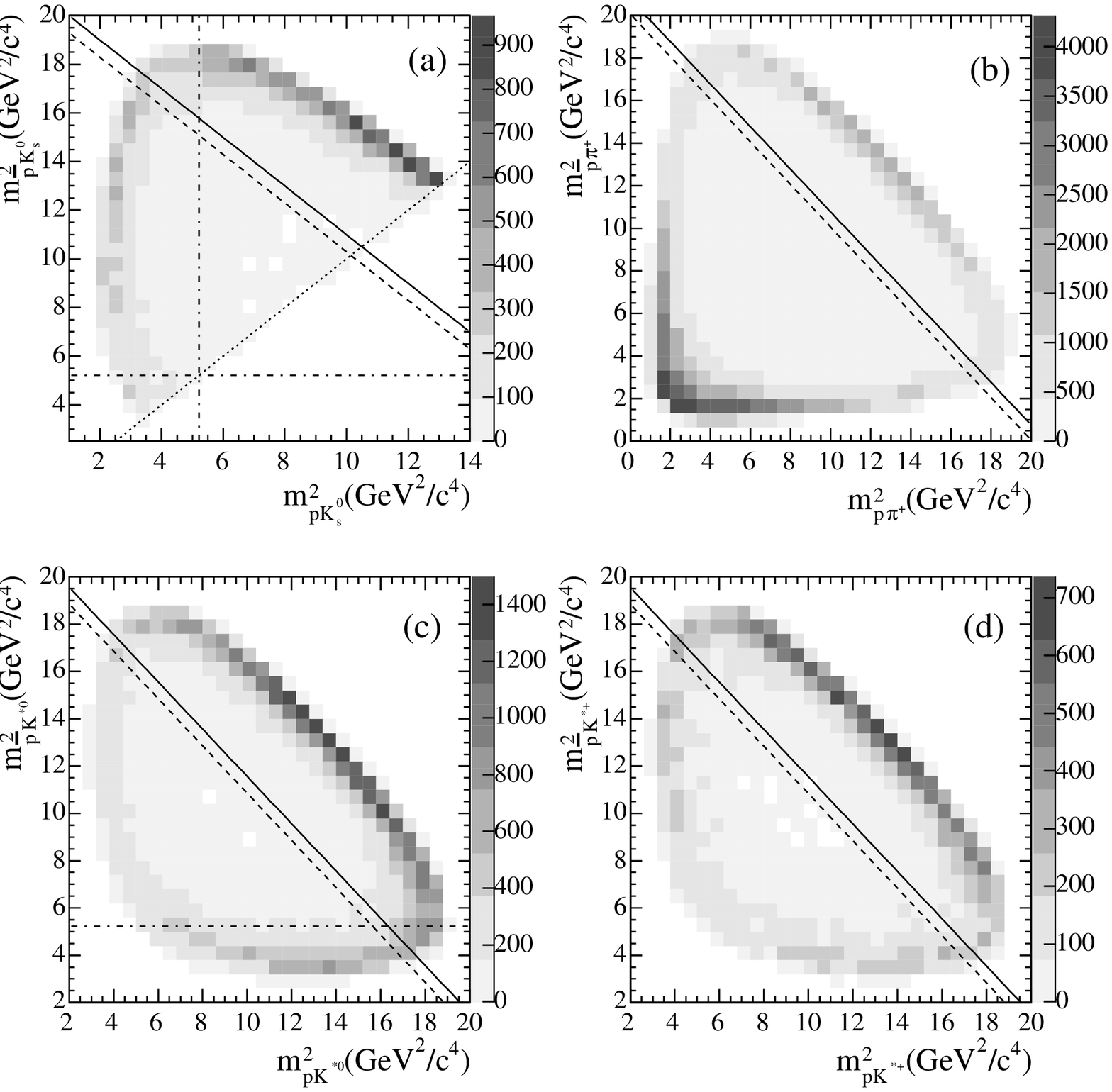}  
\caption{Dalitz plots for background in the (a) \bppkz, (b) \bpppi, (c) \bppkst and (d) \bppkstp modes 
obtained using the weighting technique described in Ref.~\cite{splot}. The expected positions of the following 
resonances are shown: $\eta_c$ in solid, $J/\psi$ in dashed and $\Lambda_c$ in dot-dashed lines.
The white areas correspond to regions with no entries.}
\label{alldalbkg}
\end{center}
\end{figure}

\section{Study of the $\boldsymbol{B\to p\overline{p}h}$ decay dynamics}

For decay dynamics studies, the maximum likelihood fit is performed using three variables 
(\mes, \DeltaE and Fisher discriminant) simultaneously over the whole Dalitz plot, 
with the exception of the 
\pppi mode where we perform the fit in two regions: \mpp$<3.6$\gevcc and \mpp$>3.6$\gevcc.

The resulting background-subtracted efficiency-corrected  
Dalitz plots for all the modes are shown in Fig.~\ref{alldal}.  
The main features of the Dalitz plots are expected to be 
the charmonium resonances (with $J/\psi\to p\overline{p}$ and  $\eta_c\to p\overline{p}$ bands most prominent), 
potential $\Lambda^+_c$ bands in \bppkz and \bppkst modes, as well as the low $p\overline{p}$ mass 
enhancements.
Figure~\ref{alldalbkg} shows the corresponding background Dalitz distributions. 
The combinatorial background events favor the edges of the
Dalitz plot, as they are dominated by the inclusion of random low-momentum tracks.

\begin{figure}
\begin{center}
\includegraphics[height=8cm,width=8.5cm]{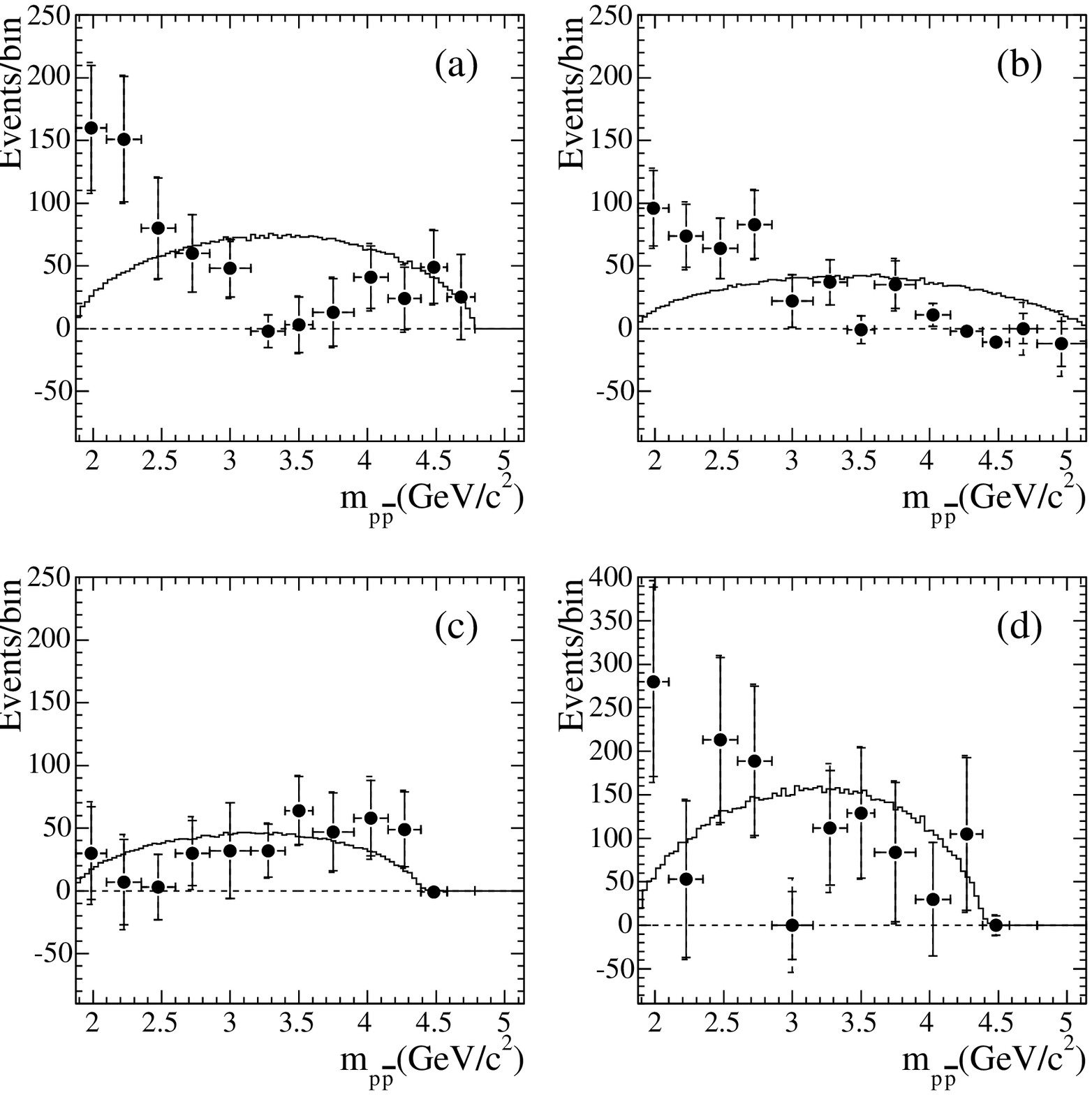}  
\caption{\mpp distribution in data (points) for (a) \bppkz, (b) \bpppi, (c) \bppkst and (d) \bppkstp modes
obtained using the weighting technique described in Ref.~\cite{splot}. The 
inner error bars show statistical uncertainty and the outer ones  
show statistical and systematic uncertainties added in quadrature. The histograms correspond to the 
relevant three-body phase-space signal Monte Carlo distributions. 
The dashed line shows zero.
The bin size is varying and is specified in Table~\ref{table-all-mpp}.
Note that the contribution from the \B decays to charmonium is removed.}
\label{all-mpp}
\end{center}
\end{figure}

Background-subtracted efficiency-corrected \mpp distributions are shown in Fig.~\ref{all-mpp} and 
are summarized in Table~\ref{table-all-mpp}. 
Although the \mpp enhancement at low mass is quite prominent in the 
\bppkz and \bpppi modes, in the case of the \bppkst and \bppkstp modes the statistics are too limited 
to draw a definite conclusion. Note that the shapes of the enhancement in \bppkz and \bppk~\cite{babar} 
are similar within the statistics of the measurements, in agreement with the theoretical predictions~\cite{dia}. 

\begin{table*}
\caption{Summary of efficiency-corrected, background-subtracted number of events in $m_{p\overline{p}}$ bins for 
\bppkz, \bpppi, \bppkst, and \bppkstp modes. Note that contributions from the \B decays to charmonium are removed.}
\label{table-all-mpp}
\begin{center}
\begin{tabular}{c++++}
\hline\hline
{\mpp range (\gevcc)}           & \multicolumn{1}{r}{\bppkz}            & \multicolumn{1}{r}{\bpppi}           & \multicolumn{1}{r}{\bppkst}                 & \multicolumn{1}{r}{\bppkstp} \\ \hline
{$1.876-2.100$}&160m50\pm14  &  96 m30\pm9  &  30m37\pm17  & 280m109\pm39 \\ 
{$2.100-2.350$}&151m50\pm13  &  74 m25\pm9  &  7 m34\pm16  & 53 m 90\pm22 \\ 
{$2.350-2.600$}&80 m40\pm7   &  64 m24\pm5  &  3 m26\pm5   & 213m 95\pm18 \\ 
{$2.600-2.850$}&60 m31\pm6   &  83 m27\pm8  &  30m26\pm12  & 189m 86\pm22 \\ 
{$2.850-3.150$}&48 m23\pm8   &  22 m21\pm3  &  32m38\pm5   & 0  m 39\pm37 \\ 
{$3.150-3.400$}&-2 m13\pm5   &  37 m18\pm3  &  32m21\pm6   & 112m 66\pm33 \\ 
{$3.400-3.600$}&3  m22\pm8   &  -1 m11\pm1  &  64m27\pm8   & 129m 75\pm14 \\ 
{$3.600-3.900$}&13 m27\pm5   &  35 m19\pm9  &  47m31\pm7   & 84 m 80\pm17 \\ 
{$3.900-4.150$}&41 m25\pm10  &  11 m 9\pm1  &  58m30\pm12  & 30 m 65\pm9  \\ 
{$4.150-4.387$}&24 m25\pm9   &  -2 m 7\pm1  &  49m30\pm5   & 105m 88\pm17 \\ 
{$4.387-4.580$}&49 m29\pm10  & -11 m 4\pm3  &  -1m6\pm1   & 0m11\pm3  \\  
{$4.580-4.782$}&25 m34\pm6   &   0 m12\pm17 & \multicolumn{1}{c}{     n/a} &  \multicolumn{1}{c}{      n/a}                 \\            
{$4.782-5.139$}&\multicolumn{1}{c}{     n/a} &      12m18\pm19 & \multicolumn{1}{c}{     n/a} & \multicolumn{1}{c}{     n/a}      \\ \hline     
\end{tabular}									       			       
\end{center}												       
\end{table*}

\begin{figure}
\begin{center}
\includegraphics[height=8cm,width=8.5cm]{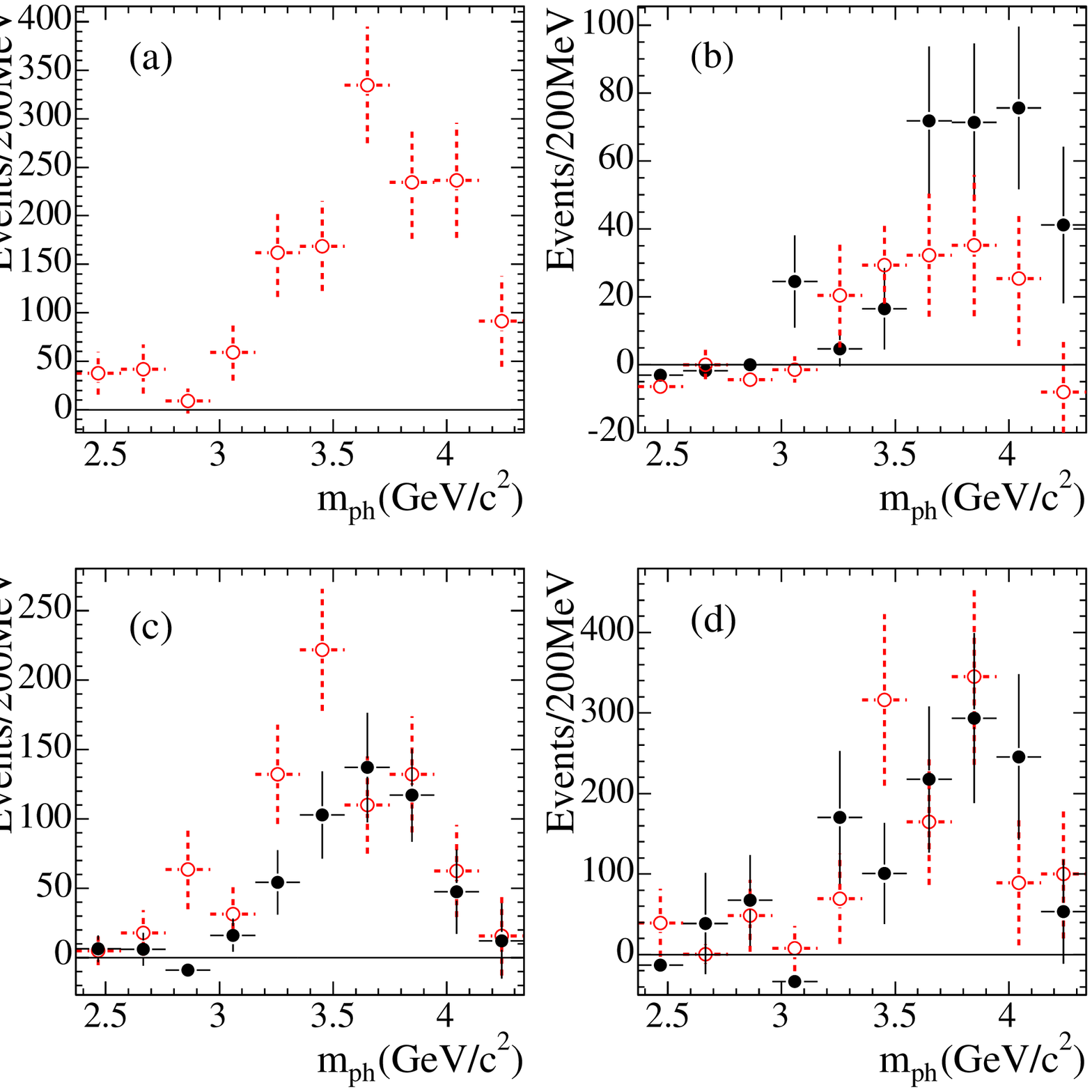}  
\caption{$m_{ph}$ distribution obtained using the weighting technique described in Ref.~\cite{splot}: 
in red (open circles) for $m_{ph}>m_{\overline{p}h}$
and in black (filled circles) for $m_{ph}<m_{\overline{p}h}$ for (a) \bppkz, (b) \bpppi, 
(c) \bppkst and (d) \bppkstp. Solid line shows zero. Only statistical error bars are shown.}
\label{all-splitdal}
\end{center}
\end{figure}

To shed light on the nature of this enhancement~\cite{Rosner}, its uniformity on the Dalitz plot has
been tested. The Dalitz plot is divided along the diagonal $m_{ph}=m_{{\overline p}h}$ line 
and each of the two halves is projected onto the nearer axis. 
If the Dalitz plot is symmetric we expect the number of events in both projections to be the same.
The corresponding background-subtracted efficiency-corrected distributions for the signal events in all the modes are 
shown in Fig.~\ref{all-splitdal}.
No asymmetry is expected to be introduced from variations in $\eps_{m_{p\overline p}}$ which is charge-symmetric 
and slowly varying with ${m_{p \overline p}}$.

In the case of \bppkz, there is no information on the flavor of the $B$  meson 
and thus this study cannot be performed. For the \bppkstp mode there seems to be 
no difference between the two halves within the available statistics. In the \bppkst mode 
there might be a marginal excess at low $m_{ph}$, which could be caused by the presence of the 
standard baryon resonances (such as $\Lambda_c^+$), while in the 
\bpppi mode there is a marginal excess of events at high $m_{ph}$ around 3.8\gevcc
in the $m_{p\pi^{+}}$ half of the Dalitz plot,
contrary to the result observed in the \bppk mode~\cite{babar}.
No quantitative theoretical description of this correlation is available at present.
Although the asymmetry in the low \mpp band shown in~\cite{babar} disfavors the possibility of  
the low mass $p\overline{p}$ enhancement originating only from the presence of a resonance below 
threshold (such as the baryonium candidate at $1835$\mevcc  recently seen by BES~\cite{BESX})
the low statistics of the current modes does not allow to derive a definite conclusion.

\section{Conclusions}

In summary, with 210$\,\mbox{fb}^{-1}$ of data, 
we report evidence for the 
$B^0\rightarrow p\overline{p}K^{*0}$ decay with a branching fraction 
 (1.5$\pm$0.5(stat)$\pm$0.4(syst))$\times 10^{-6}$, 
and provide improved measurements of branching fractions of the other 
\bpph modes, where $h=\pi^+,\,K^0$ and $K^{*+}$. 
One key observation is that the pattern of decays for $B\rightarrow p\overline{p}h$
differs from that found for $B\to\pi^0 h$ and $\B\to\rho^0h$.
We also identify decays of the type $B\rightarrow X_{c\overline{c}}h\rightarrow p\overline{p} h$, 
where $h=K_S^0,\,K^{*0}$ and $K^{*+}$, and $X_{c\overline{c}}=\eta_c$ and $J/\psi$.
In particular, we report evidence for the $B^+\rightarrow \eta_cK^{*+}$ decay 
with the branching fraction of 
${\cal B}(B^+\rightarrow\eta_cK^{*+})\times{\cal B}(\eta_c\to p\overline{p})=$
($1.57^{+0.56}_{-0.45}$(stat)$^{+0.46}_{-0.36}$(syst))$\times 10^{-6}$.
We confirm the earlier observation of the $B^0\to \Lambda^+_c\overline{p}$ decay~\cite{2bodyex2} and 
report measurements of the charge asymmetry consistent with zero 
in the \bpppi, \bppkst and \bppkstp modes.
No evidence is found for the pentaquark candidate $\Theta^{+}$
in the mass range 1.52 to 1.55\gevcc, decaying into $p\KS$, 
or the glueball candidate $f_J(2220)$ in 
the mass range $2.2<\mpp<2.4$\gevcc, 
and branching fraction limits are established at 
the $10^{-7}$ level.

% Standard acknowledgments paragraph; must always be included.
\section{ACKNOWLEDGMENTS}
\label{sec:acknowledgments}
We are grateful for the 
extraordinary contributions of our \pep2\ colleagues in
achieving the excellent luminosity and machine conditions
that have made this work possible.
The success of this project also relies critically on the 
expertise and dedication of the computing organizations that 
support \babar.
The collaborating institutions wish to thank 
SLAC for its support and the kind hospitality extended to them. 
This work is supported by the
US Department of Energy
and National Science Foundation, the
Natural Sciences and Engineering Research Council (Canada),
the Commissariat \`a l'Energie Atomique and
Institut National de Physique Nucl\'eaire et de Physique des Particules
(France), the
Bundesministerium f\"ur Bildung und Forschung and
Deutsche Forschungsgemeinschaft
(Germany), the
Istituto Nazionale di Fisica Nucleare (Italy),
the Foundation for Fundamental Research on Matter (The Netherlands),
the Research Council of Norway, the
Ministry of Science and Technology of the Russian Federation, 
Ministerio de Educaci\'on y Ciencia (Spain), and the
Science and Technology Facilities Council (United Kingdom).
Individuals have received support from 
the Marie-Curie IEF program (European Union) and
the A. P. Sloan Foundation.


\begin{thebibliography}{99}
\bibitem{chic0other} There are $1.70\pm2.42$, $1.66\pm 1.74$, $-0.39\pm0.06$ and $-0.86\pm 0.70$ events in the $\chi_{c0}$ region for each of the modes shown in Table~\ref{bfres} respectively.
\end{thebibliography}
\end{document}